\documentclass[10pt,twocolumn]{IEEEtran}

\usepackage{graphicx}
\usepackage{amsmath}
\usepackage{amssymb}
\usepackage[caption=false]{subfig}
\usepackage[noadjust]{cite}
\usepackage{float}
\usepackage{algorithm}
\usepackage{bibentry}
\usepackage{balance}
\usepackage{algorithm}
\usepackage{algorithmicx}
\usepackage{algpseudocode}
\usepackage{xcolor}
\usepackage{subfig}
\usepackage{array}
\usepackage{wrapfig}
\usepackage{makecell}

\usepackage{url}

\graphicspath{{img/}}

\begin{document}

\bstctlcite{IEEEexample:BSTcontrol}

\title{Kriging-Based 3-D Spectrum Awareness for Radio Dynamic Zones Using Aerial Spectrum Sensors
\thanks{This work is supported in part by the NSF PAWR award CNS-1939334 and its associated supplement for studying National Radio Dynamic Zones (NRDZs). The authors would like to thank Wireless Research Center for measuring antenna patterns by using an anechoic chamber. The datasets and post-processing scripts for obtaining the results in this manuscript are publicly accessible at~\cite{IEEEDataPort}.}\thanks{S. J. Maeng, Ozgur Ozdemir, \.{I}. G\"{u}ven\c{c}, and Mihail L. Sichitiu are with the Department of Electrical and Computer Engineering, North Carolina State University, Raleigh, NC 27606 USA (e-mail: smaeng@ncsu.edu; oozdemi@ncsu.edu;  iguvenc@ncsu.edu; mlsichit@ncsu.edu).}}

\author{\IEEEauthorblockN{Sung Joon Maeng, Ozgur Ozdemir, \textit{Member, IEEE}, \.{I}smail G\"{u}ven\c{c}, \textit{Fellow, IEEE}, and Mihail L. Sichitiu, \textit{Member, IEEE}}}

\maketitle

\begin{abstract}
Radio dynamic zones (RDZs) are geographical areas within which dedicated spectrum resources are monitored and controlled to enable the development and testing of new spectrum technologies. Real-time spectrum awareness within an RDZ is critical for preventing interference with nearby incumbent users of the spectrum. In this paper, we consider a 3D RDZ scenario and propose to use unmanned aerial vehicles (UAVs) equipped with spectrum sensors to create and maintain a 3D radio map of received signal power from different sources within the RDZ. In particular, we introduce a 3D Kriging interpolation technique that uses realistic 3D correlation models of the signal power extracted from extensive measurements carried out at the NSF AERPAW platform. 
%based on Kriging interpolation of signal power. Our experiment was conducted at the Aerial Experimentation and Research Platform for Advanced Wireless (AERPAW) test site where an unmanned aerial vehicle (UAV) collected signal from an LTE base station (BS) tower. 
% Based on extensive real-world measurements carried out at the NSF AERPAW Platform at NC State University. 
%We performed post-processing on the collected in-phase and quadrature (I/Q) sample data,
Using C-Band signal measurements by a UAV at altitudes between 30~m-110~m, we first develop realistic propagation models on air-to-ground path loss, shadowing, spatial correlation, and semi-variogram, while taking into account the knowledge of antenna radiation patterns and ground reflection. Subsequently, we generate a 3D radio map of a signal source within the RDZ using the Kriging interpolation and evaluate its sensitivity to the number of measurements used and their spatial distribution. Our results show that the proposed 3D Kriging interpolation technique provides significantly better radio maps when compared with an approach that assumes perfect knowledge of path loss.

\end{abstract}


\section{Introduction}\label{sec:intro}
As the demand for advanced wireless communication services continues to grow, efficient use of spectrum resources is becoming increasingly vital for future wireless technologies. Therefore, the development, testing, and evaluation of effective mechanisms to improve spectrum efficiency and sharing have become imperative. Although there is a considerable body of literature that examines and analyzes spectrum sharing using theoretical models and simulations, there is a clear need to assess these approaches in real-world deployment scenarios, taking into account realistic propagation conditions.

In this particular context, the concept of radio dynamic zones (RDZs) emerges as a new concept~\cite{zheleva2023radio,maeng2022national,tschimben2023testbed}, where geographical areas with dedicated spectrum resources are effectively managed and controlled in real-time to test new wireless innovations. This management is achieved through the sensing of signals entering and leaving the zone~\cite{NSF_SIINRDZ_Solicitation}. RDZs serve as testing grounds for novel spectrum sharing concepts and emerging technologies aimed at improving spectrum efficiency within specific deployment scenarios. In RDZs, it becomes crucial to ensure minimal or no interference to existing incumbent users of the spectrum. Therefore, monitoring of signal leakage to passive or active receivers outside the RDZ becomes necessary. This requires installation and deployment of sensors within the RDZ. Monitoring scope can include both terrestrial areas and airspace, e.g., for coexistence with unmanned aerial vehicles (UAVs) and satellites. By monitoring and modeling the interference levels experienced by passive receivers in these aerial scenarios, more efficient spectrum sharing can be achieved.

The use of radio environment maps (REMs)~\cite{8057286} presents an effective approach for constructing dynamic interference maps within an RDZ, which can be generated for each location and frequency of interest. These radio maps are generated by collecting signal power data from deployed sensors and incorporating their corresponding location information. However, it is often impractical to position sensors throughout the entire RDZ area. Instead, signal power at unknown locations can be predicted using signal processing techniques like Kriging~\cite{6685772}, based on measurements from nearby sparsely deployed sensors. Kriging takes advantage of the spatial correlation between different locations to optimize the prediction of signal power. By employing Kriging, we can efficiently interpolate and generate a radio map of signal power using sparsely measured datasets from the sensors.

In the existing literature, several studies have focused on modeling the spatial correlation of shadowing in received signals~\cite{991146,5590312}, with experimental measurements provided in~\cite{gudmundson1991correlation,6882778}. The application of Kriging for generating radio maps of signal power has been validated using both simulated and real datasets~\cite{7542153}. The potential of Kriging for spectrum monitoring and interference management has been explored in ~\cite{7817747}, while \cite{9316892} extends Kriging interpolation to spectrum interpolation and analyzes it using measurement datasets. For ground-to-UAV communications in suburban environments, path loss and shadowing have been modeled based on measurement datasets~\cite{8048502,4483593}. Additionally, the spatial correlation along the linear trajectory of a UAV has been investigated~\cite{6492100}. In our recent works, we introduce the RDZ concept and discuss its features and requirements~\cite{maeng2022national}. Furthermore, we propose a leakage sensing algorithm using Kriging in the two-dimensional (2D) plane of the RDZ~\cite{maeng2021out}. Notably, to the best of our knowledge, the literature does not address the use of Kriging to obtain a three-dimensional (3D) aerial radio map based on measurements obtained from unmanned aerial vehicles (UAVs).

In this paper, we propose to develop and use a 3D radio map to effectively sense signal leakage from an RDZ to the receivers outside of the RDZ. We employ a UAV as a mobile aerial sensor, collecting signal power measurements from distinct receivers within the RDZ. %along a trajectory that sweeps the experimental site. 
The 3D interpolation of the collected signal power is performed using the Kriging technique. The proposed method is thoroughly analyzed and validated through a measurement campaign. The main contributions of this paper can be summarized as follows: 
\begin{itemize}
    \item Modeling 3D Radio Propagation: Considering a 3D spectrum sensing scenario, we develop and analyze a   path loss model that accounts for spatially correlated shadowing, two-ray wireless propagation, and \emph{measured} antenna radiation patterns to accurately model 3D radio propagation. We integrate 3D antenna measurements obtained in an anechoic chamber and study improvements in model accuracy when compared to using dipole and omnidirectional antenna patterns.
    % \item Analysis of Elevation-Dependent Antenna Radiation Pattern: We utilize the measurement data to analyze the impact of elevation-dependent antenna radiation patterns on radio propagation. Furthermore, we develop a model to characterize and understand this behavior.
    \item Semi-Variogram Based Kriging Interpolation: We introduce a novel method for Kriging interpolation specifically designed for 3D spectrum monitoring. This approach leverages a semi-variogram technique to achieve accurate and efficient interpolation across a 3D volume using a limited set of measurements.
    \item Comparison with Measurement Data: We evaluate and compare the accuracy of our proposed 3D propagation models with the measurement data collected using software-defined radios (SDRs) at various UAV altitudes. This analysis provides valuable insights into the performance and reliability of the proposed approach.
\end{itemize}

The rest of this paper is organized as follows. In Section~\ref{sec:system}, we present the system model for 3-D spectrum sensing, radio propagation, and spatial correlation in an RDZ, while in Section~\ref{sec:3D_Kriging}, we introduce the Kriging-based signal interpolation method for generating a 3D radio map. In Section~\ref{sec:meas_camp}, we describe the details of our measurement campaigns for obtaining I/Q signal samples at a UAV from an LTE-based signal source on the ground, and our measurements in an anechoic chamber for characterizing the antenna radiation patterns. In Section~\ref{sec:A2G_ana}, we analyze the effectiveness of the proposed 3D path-loss models in predicting the received signal power at different UAV altitudes and locations.  We present numerical results on Kriging-based 3D radio map interpolation for various scenarios in Section~\ref{sec:sim_Kriging} and the last section concludes the paper.

\section{System Model} \label{sec:system}
In this section, we present the models utilized for spectrum sensing within an RDZ. Specifically, we consider a scenario where an aerial spectrum sensor traverses the area and captures received signals from a base station (BS). Radio propagation, correlation, and antenna radiation pattern models are also presented.

\subsection{3-D Spectrum Sensing with an Aerial Mobile Sensor}
\begin{figure}[t]
	\centering
	\includegraphics[width=0.48\textwidth]{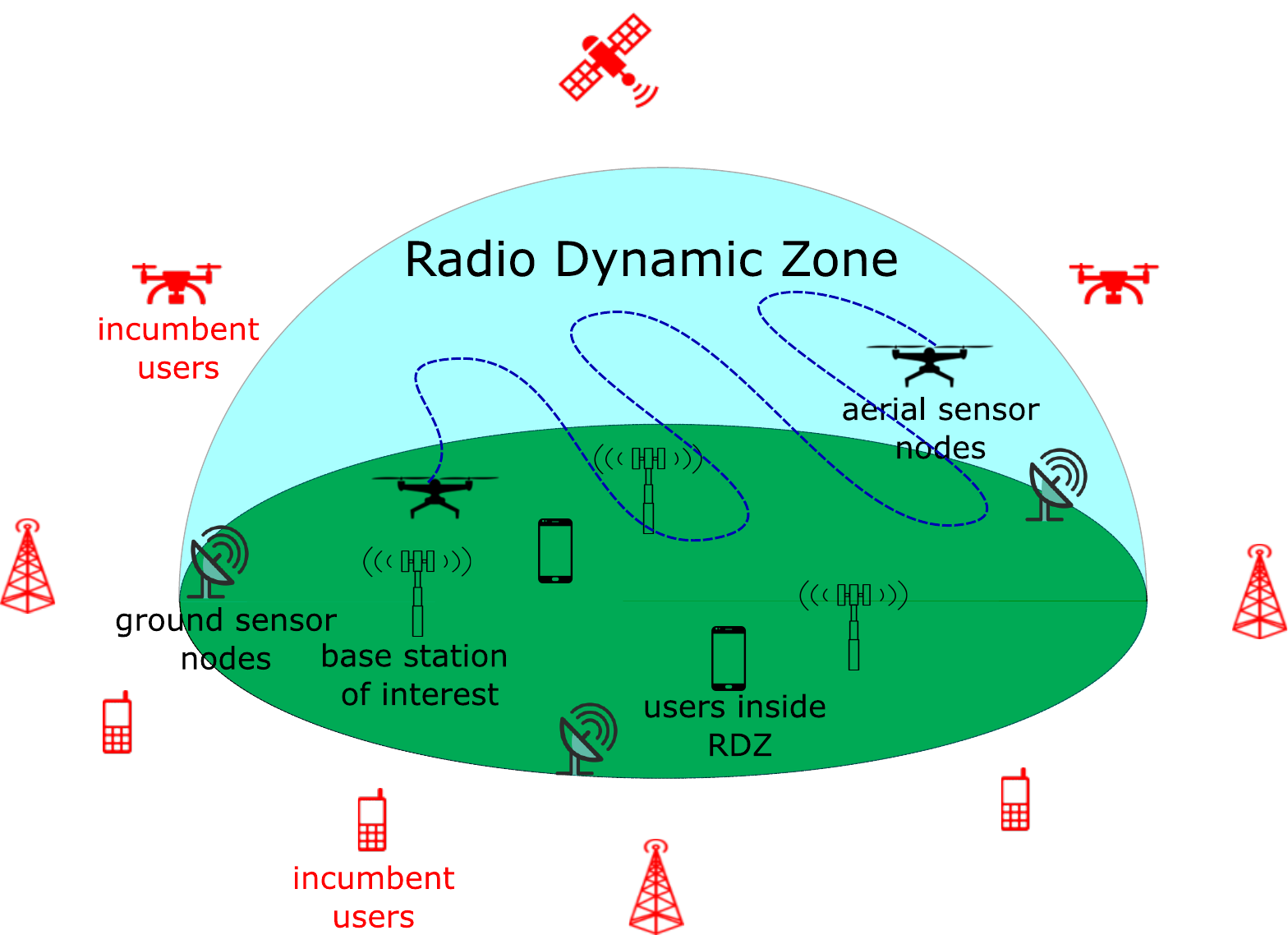}
	\caption{Illustration of an RDZ with aerial and ground sensors and users.}\label{fig:RDZ_illu}
\end{figure}

An RDZ should protect incumbent users outside of the zone by controlling and managing interference signals radiating from inside the zone. The incumbent users may include smart devices and aerial vehicles, as well as sensitive scientific passive receivers such as satellites and ground-based radio astronomy receivers in radio quiet zones (RQZs)~\cite{NRDZ_vs_NRQZ}. Our envisioned RDZ concept is illustrated in Fig.~\ref{fig:RDZ_illu}. The real-time spectrum sensing within the boundary of the RDZs is conducted by deployed fixed / mobile ground and aerial sensor nodes, which is an essential technique to manage dynamic spectrum usage. The UAV moves across the RDZ space along a multi-altitude trajectory, capturing signal data throughout.

% Our works in this paper mainly focuses on the study of real-time signal sensing on the volume of space to monitor the signal leakage of RDZ. UAVs as mobile aerial nodes collect signal power following the predefined trajectories. After that, the RDZ system generates the radio map of signal power surrounding the RDZ space by interpolating the collected dataset from the aerial nodes. A system model of collecting the signal transmitted from a single BS at the aerial node is described in Fig.~\ref{fig:sensing_illu}. The UAV sweeps across the RDZ space following the multi-altitude trajectory.

% RDZs are designed to protect incumbent users located outside of the zones by effectively controlling and managing interference signals originating from within the zones. These incumbent users encompass not only smart devices and aerial vehicles but also sensitive scientific passive receivers like satellites and receivers in radio quiet zones (RQZs)~\cite{NRDZ_vs_NRQZ}. The concept of RDZs is visually depicted in Fig.~\ref{fig:RDZ_illu}. Real-time spectrum sensing is a fundamental technique employed within the boundaries of RDZs, which relies on fixed and mobile ground as well as aerial sensor nodes. This sensing approach is crucial for dynamically managing spectrum usage.

This paper primarily focuses on the study of real-time signal sensing in the volume of space to monitor the signal leakage from RDZs. Mobile aerial nodes, in the form of UAVs, collect signal power data as they follow predefined trajectories. Subsequently, the RDZ system leverages the collected dataset from the aerial nodes to generate a radio map depicting the signal power surrounding the RDZ space. The interpolation of this dataset facilitates the construction of a comprehensive representation of signal power distribution. 
%Fig.~\ref{fig:sensing_illu} illustrates a system model where the aerial node captures the signals transmitted from a BS \textcolor{red}{of interest}. 

% \begin{figure}[t]
% 	\centering
% 	\includegraphics[width=0.39\textwidth]{3D_sensing_illu.pdf}
% 	\caption{Illustration of signal sensing by an aerial mobile sensor node.}\label{fig:sensing_illu}
% \end{figure}

\subsection{Radio Propagation Model}
The location of a BS and a UAV can be represented by
\begin{align}
    \mathbf{l}^{\rm bs}&=(\psi^{\rm bs},\omega^{\rm bs},h^{\rm bs}),\;\mathbf{l}^{\rm uav}(t)=(\psi^{\rm uav},\omega^{\rm uav},h^{\rm uav}),
\end{align}
where $\psi$, $\omega$, and $h$ denote the latitude, longitude, and altitude of the location. Note that although the location can be generally represented by \textit{x, y, z} in 3D Cartesian coordinates, we express it by latitude, longitude, and altitude to use the information given by GPS sensors. The time-varying location of a UAV is given by $\mathbf{l}^{\rm uav}(t)$. The horizontal distance and the vertical distance between a BS and a UAV can be expressed as~\cite{chopde2013landmark}
\begin{align}
    d_{\rm h}(\mathbf{l}^{\rm bs},\mathbf{l}^{\rm uav})&=\arccos\left(\sin\psi^{\rm uav}\sin\psi^{\rm bs}\right.\nonumber\\
    &\left.+\cos\psi^{\rm uav}\cos\psi^{\rm bs}\cos(\omega^{\rm bs}-\omega^{\rm uav})\right)\times A,\\
    d_{\rm v}(\mathbf{l}^{\rm bs},\mathbf{l}^{\rm uav})&=|h^{\rm bs}-h^{\rm uav}|,
\end{align}
where $A$ is the radius of the earth ($\approx 6378137$ m). Then, the 3D distance between a BS and a UAV is given by
\begin{align}
    d_{\rm 3D}(\mathbf{l}^{\rm bs},\mathbf{l}^{\rm uav})&=\sqrt{d_{\rm h}(l^{\rm bs},l^{\rm uav})^2+d_{\rm v}(l^{\rm bs},l^{\rm uav})^2}.
\end{align}
Next, the elevation angle between a BS and a UAV can be expressed as
\begin{align}
    \theta_l&=\tan^{-1}\left(\frac{ d_{\rm v}}{d_{\rm h}}\right).
\end{align}

\begin{figure}[t]
	\centering
	\includegraphics[width=0.48\textwidth]{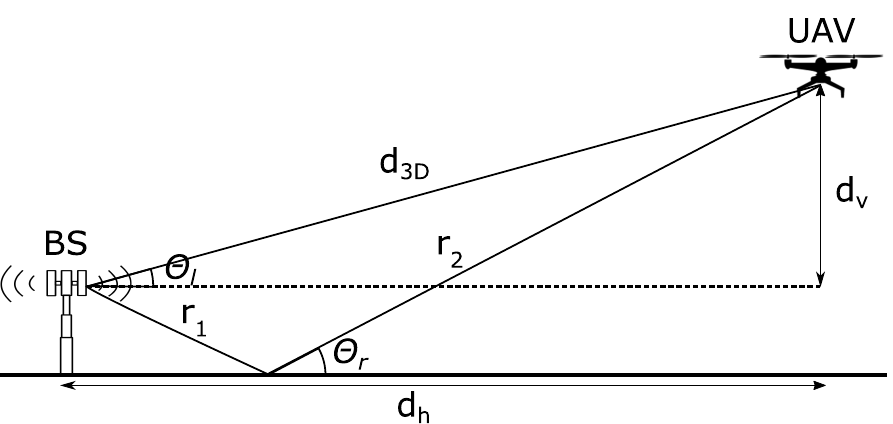}
	\caption{The illustration of the two-ray ground reflection model.}\label{fig:two_ray_illu}
\end{figure}

To develop a propagation model, we make use of a first-order approximation and consider the rural environment in which we collect measurements. In this scenario, we employ the two-ray ground reflection model to represent the path loss between a BS and a UAV. This model accounts for a line-of-sight (LoS) path as well as a strong ground reflection path, both contributing to the received signal as the two dominant paths in an open area such as a rural environment. The path loss characterized by the two-ray ground reflection model can be expressed as follows~\cite[Chapter 2]{jakes1994microwave}:
\begin{align}\label{eq:PL_two}
    &\mathsf{PL}_{\rm twm}(\mathbf{l}^{\rm bs},\mathbf{l}^{\rm uav})=\left(\frac{\lambda}{4\pi}\right)^2\bigg|\underbrace{\frac{\sqrt{\mathsf{G}_{\rm bs}(\phi_l,\theta_l)\mathsf{G}_{\rm uav}(\phi_l,\theta_l)}}{d_{\rm 3D}}}_{\text{LoS signal}}\nonumber\\
    &+\underbrace{\frac{\Gamma(\theta_r)\sqrt{\mathsf{G}_{\rm bs}(\phi_r,\theta_r)\mathsf{G}_{\rm uav}(\phi_r,\theta_r)}e^{-j\Delta\tau}}{r_1+r_2}}_{\text{ground reflected signal}}\bigg|^2,
\end{align}
where $\mathsf{G}_{\rm bs}(\phi,\theta)$, $\mathsf{G}_{\rm uav}(\phi,\theta)$, $\lambda$, $\phi$ denote the antenna gain of a BS, antenna gain of a UAV, wave-length, and azimuth angle, respectively, $\theta_r=\tan^{-1}\left(\frac{h^{\rm bs}+h^{\rm uav}}{d_{\rm h}}\right)$ represents ground reflection angle, and $\Delta\tau=\frac{2\pi(r_1+r_2-d_{\rm 3D})}{\lambda}$ indicates the phase difference between two paths. The distance and the angle parameters in the two-ray ground reflection model are illustrated in Fig.~\ref{fig:two_ray_illu}. The ground reflection coefficient with the vertically polarized signal is given by
\begin{align}
    \Gamma(\theta_r)&=\frac{\varepsilon_0\sin\theta_r-\sqrt{\varepsilon_0-\cos^2\theta_r}}{\varepsilon_0\sin\theta_r+\sqrt{\varepsilon_0-\cos^2\theta_r}},
\end{align}
where $\varepsilon_0$ is the relative permittivity of the ground and the value depends on the type of the ground. Two signal components in \eqref{eq:PL_two} are received and combined with a phase difference. If we only consider the first LoS term in the path loss, we can obtain the free-space path loss model, given as
\begin{align}\label{eq:PL_fs}
    &\mathsf{PL}_{\rm fs}=\left(\frac{\lambda}{4\pi}\right)^2\left|\frac{\sqrt{\mathsf{G}_{\rm bs}(\theta_l)\mathsf{G}_{\rm uav}(\theta_l)}}{d_{\rm 3D}}\right|^2.
\end{align}

Using \eqref{eq:PL_two}, the received signal power of a UAV in dB scale can be expressed as
\begin{align}\label{eq:received}
    r&=\mathsf{P}_{\rm Tx}-\mathsf{PL}_{\rm twm}^{(\rm dB)}+w,
\end{align}
where $\mathsf{P}_{\rm Tx}$, $w$ denote transmit power and shadowing component, respectively. Note that the path loss term in \eqref{eq:received} is converted to dB scale. The shadowing term generally follows a lognormal distribution and is modeled by a zero-mean Gaussian process with a spatial covariance~\cite{991146}. The correlation between received signals at two different locations is generally characterized by the function of the distance between those locations. Note that we do not take into account small-scale fading in the received signal since we assume that the effect is eliminated by averaging the samples within the proper time interval~\cite{gudmundson1991correlation}. 

\subsection{Spatial Correlation Model of Received Signal}
In this section, we focus on describing the correlation function between the received signals at different locations of a UAV. Since the spatial correlation primarily depends on the shadowing component ($w$) in the received signal in \eqref{eq:received}, we can capture the correlation between received signals ($r$) using the correlation between the shadowing components without loss of generality. It is well-known that the correlation between two different locations is characterized by a function of their physical distance. Typically, this correlation exponentially attenuates as the physical distance between the locations increases~\cite{gudmundson1991correlation}. However, most existing works in the literature primarily focus on terrestrial networks and do not fully consider 3D topologies. Due to this limitation, the spatial correlation between two locations with different vertical positions (heights) has not been extensively studied to our best knowledge. Considering the unique characteristics of UAV-based scenarios, where altitude plays a crucial role, it becomes essential to investigate and understand the spatial correlation between locations at different vertical positions. This exploration will allow for a more comprehensive modeling of the correlation in 3D scenarios, considering the impact of vertical distance in addition to horizontal distance.

In our work, we first model the spatial correlation as a function of the vertical distance ($d_{\rm v}$) as well as the horizontal distance ($d_{\rm h}$). Then, we define the correlation function between 3D locations as a function of both the vertical distance and the horizontal distance. The spatial correlation between two different locations of a UAV, i.e., between $l^{\rm uav}_i$ and $l^{\rm uav}_j$, can be expressed as
\begin{align}\label{eq:corr_def}
    R(l^{\rm uav}_i,l^{\rm uav}_j)=R(d_{\rm v},d_{\rm h})=\frac{\mathbb{E}\left[w(l^{\rm uav}_i)w(l^{\rm uav}_j)\right]}{\sigma_{w}^2},
\end{align}
where $\sigma_{w}^2$ is the variance of shadowing. Once again, the proposed correlation is the function of both the vertical distance and  the horizontal distance.

\subsection{Antenna Radiation Model}
The antenna gain effect of a transmitter and a receiver in the received signal is captured in the path loss model in \eqref{eq:PL_two}, using $\mathsf{G}_{\rm bs}(\phi,\theta)$, $\mathsf{G}_{\rm uav}(\phi,\theta)$. In typical terrestrial communications, the antenna gain is simply modeled by a constant gain. This is due to the fact that a dipole antenna is usually characterized as an omni-directional antenna radiation pattern in the azimuth angle domain, or sectored directional antennas make the antenna pattern mostly uniform in the azimuth angle domain. However, air-to-ground communications require considering the variation of the antenna gain in the elevation angle domain. The antenna pattern in the elevation domain is typically far from being uniform and therefore we should consider the elevation angle-dependent radiation pattern in modeling the antenna gain.

% In the path loss model presented in \eqref{eq:PL_two}, the antenna gain effect of both the transmitter and receiver is accounted for using $\mathsf{G}_{\rm bs}(\theta)$, $\mathsf{G}_{\rm uav}(\theta)$, respectively. In typical terrestrial communications, the antenna gain is often simplified and modeled as a constant gain. This simplification is primarily because dipole antennas are commonly characterized as having an omnidirectional radiation pattern in the azimuth angle domain, or sectored directional antennas are used to achieve a uniform antenna pattern in the azimuth angle domain. However, in air-to-ground communications scenarios, it is crucial to consider the variation of the antenna gain in the elevation angle domain. Unlike in terrestrial communications, the antenna pattern in the elevation angle domain is not uniform. Therefore, it becomes necessary to incorporate an elevation angle-dependent antenna radiation pattern when modeling the antenna gain. By considering the variation of the antenna gain in the elevation angle domain, a more accurate representation of the antenna performance in air-to-ground communications can be achieved.

\section{3D Radio Map Interpolation using Kriging}\label{sec:3D_Kriging}
In this section, we introduce an efficient radio map interpolation technique using Kriging~\cite{7817747}. This method utilizes measurement data obtained from sparsely deployed spectrum sensors within an RDZ. The interpolation process allows us to estimate signal values at unsampled locations based on the available measurements. We first introduce how to calculate a semi-variogram, and subsequently, introduce our Kriging based interpolation approach for 3D RDZ scenarios. Different than the existing Kriging techniques in the literature, we consider the 3D geometry in spatial correlation with a portable aerial sensor, which enables us to interpolate the radio map in a 3D volume.

\subsection{Semi-variogram}\label{sec:semi-var}
In geostatistics, the semi-variogram represents the degree of spatial dependency on different locations which is utilized in Kriging interpolation. The semi-variogram between a UAV's locations $l^{\rm uav}_i$, $l^{\rm uav}_j$ is defined as
\begin{align}\label{eq:semi_var_def}
    \gamma(l^{\rm uav}_i,l^{\rm uav}_j)=\frac{1}{2}\text{var}\left(r(l^{\rm uav}_i)-r(l^{\rm uav}_j)\right).
\end{align}
If the covariance function of a stationary process exists, we can obtain the semi-variogram from the spatial correlation in \eqref{eq:corr_def} as follows for our considered 3D RDZ scenario~\cite{cressie2015statistics}:
\begin{align}\label{eq:var2cor}
    &\gamma(l^{\rm uav}_i,l^{\rm uav}_j)\nonumber\\
    &=\frac{\sigma_{w}^2}{2}\left(R(l^{\rm uav}_i,l^{\rm uav}_i)+R(l^{\rm uav}_j,l^{\rm uav}_j)-2R(l^{\rm uav}_i,l^{\rm uav}_j)\right)\nonumber\\
    &=\sigma_{w}^2\left(1-R(l^{\rm uav}_i,l^{\rm uav}_j)\right)=\sigma_{w}^2\left(1-R(d_{\rm v},d_{\rm h})\right),
\end{align}
where $\sigma_{w}^2$ captures the variance of the shadowing term $w$ in \eqref{eq:received} as defined earlier, and $R(l^{\rm uav}_i,l^{\rm uav}_i)$ is as defined in \eqref{eq:corr_def}. We assume that $\sigma_{w}^2$ is constant at given set of locations while deriving \eqref{eq:var2cor}.

\subsection{Kriging Interpolation}\label{sec:Kriging_ana}
The ordinary Kriging is the optimal prediction method in squared-error loss from the observed data at known spatial locations where the error of the spatial prediction of an unknown location is minimized~\cite{cressie2015statistics}. It interpolates the signal strength of the arbitrary locations by using the linear combination of the signal strength of the nearby locations. The ordinary Kriging problem can be formulated as follows~\cite{7817747}:
\begin{IEEEeqnarray}{rl}
    \min_{\mu_{1},\dots,\mu_{M}}
    &\quad \mathbb{E}\left[\left(\hat{r}(\mathbf{l}^{\rm uav}_0)-r(\mathbf{l}^{\rm uav}_0)\right)^2\right],\\
    \text{s.t.}
    &\quad \hat{r}(\mathbf{l}^{\rm uav}_0)=\sum_{i=1}^{M}\mu_{i}r(\mathbf{l}^{\rm uav}_i)\IEEEyessubnumber,\\
    &\quad \sum_{i=1}^{M}\mu_{i}=1 \IEEEyessubnumber,
\end{IEEEeqnarray}
where $l^{\rm uav}_0$ is a location to predict an unknown parameter, $\mu_i~(i=1,\cdots,M)$ are weighting parameters and $M$ indicates the number of nearby measured samples to use. 

The above problem can be solved by following steps~\cite{7817747}. First, we convert the original problem to an equivalent Lagrange expression:
\begin{align}\label{eq:prob_Lagr}
    \min_{\mu_{1},\dots,\mu_{M}}&\mathbb{E}\left[\left(r(l^{\rm uav}_0)-\sum_{i=1}^{M}\mu_{i}r(l^{\rm uav}_i)\right)^2\right]-\kappa\left(\sum_{i=1}^{M}\mu_{i}-1\right),
\end{align}
where $\kappa$ denotes the Lagrange multiplier. After a few mathematical steps, the objective function in \eqref{eq:prob_Lagr} can be reformulated as
\begin{align}\label{eq:obj_rewri}
    &\sigma^2_{w}+2\sum_{i=1}^{M}\mu_{i}\gamma(l^{\rm uav}_0,l^{\rm uav}_i)-\sum_{i=1}^{M}\sum_{j=1}^{M}\mu_{i}\mu_{j}\gamma(l^{\rm uav}_i,l^{\rm uav}_j)\nonumber\\
    &-\kappa\left(\sum_{i=1}^{M}\mu_{i}-1\right),
\end{align}
where $\gamma(l^{\rm uav}_i,l^{\rm uav}_j)$ is as defined in \eqref{eq:semi_var_def}. Finally, we can find the optimal solution that minimizes the objective function by the first derivative of \eqref{eq:obj_rewri} with respect to $\mu_1,\dots,\mu_M$, which is given by
\begin{align}\label{eq:first_der}
    \sum_{j=1}^{M}\mu_j\gamma(l^{\rm uav}_i,l^{\rm uav}_j)-\gamma(l^{\rm uav}_0,l^{\rm uav}_i)+\kappa'=0.
\end{align}

We can also express \eqref{eq:first_der} as a linear matrix equation as:
\begin{align}\label{eq:matrix_form}
    &\left[ \begin{array}{c c c c}
        \gamma(l^{\rm uav}_1,l^{\rm uav}_1) & \cdots & \gamma(l^{\rm uav}_1,l^{\rm uav}_M) & 1\\
       \gamma(l^{\rm uav}_2,l^{\rm uav}_1) & \cdots & \gamma(l^{\rm uav}_2,l^{\rm uav}_M) & 1\\
        \vdots & \vdots & \vdots & \vdots\\
        \gamma(l^{\rm uav}_M,l^{\rm uav}_1) & \cdots & \gamma(l^{\rm uav}_M,l^{\rm uav}_M) & 1\\
        1 & \cdots & 1 & 0\\
    \end{array}\right]
    \left[ \begin{array}{c}
        \mu_1\\\mu_2\\ \vdots\\ \mu_M\\ \kappa'
    \end{array}\right]\nonumber\\
    &=\left[ \begin{array}{c}
        \gamma(l^{\rm uav}_0,l^{\rm uav}_1)\\\gamma(l^{\rm uav}_0,l^{\rm uav}_2)\\ \vdots\\ \gamma(l^{\rm uav}_0,l^{\rm uav}_M)\\1
    \end{array}\right].
\end{align}
Then, we can easily obtain the optimal $\mu_1^{\star},\dots,\mu_{M}^{\star}$ from \eqref{eq:matrix_form} and interpolate the received signal powers of unknown location $l^{\rm uav}_0$ by
\begin{align}\label{eq:Kriging_sol}
    \hat{r}(l^{\rm uav}_0)=\sum_{i=1}^{M}\mu_{i}^{\star}r(l^{\rm uav}_i).
\end{align}
Note that accurate characterization of the 3D semi-variogram in \eqref{eq:semi_var_def} is critical for the interpolation in \eqref{eq:Kriging_sol}. The next section describes our measurements that will be used to characterize the 3D semi-variogram.

\begin{figure}[t]
	\centering
	\subfloat[Experiment site]{\includegraphics[width=0.48\textwidth]{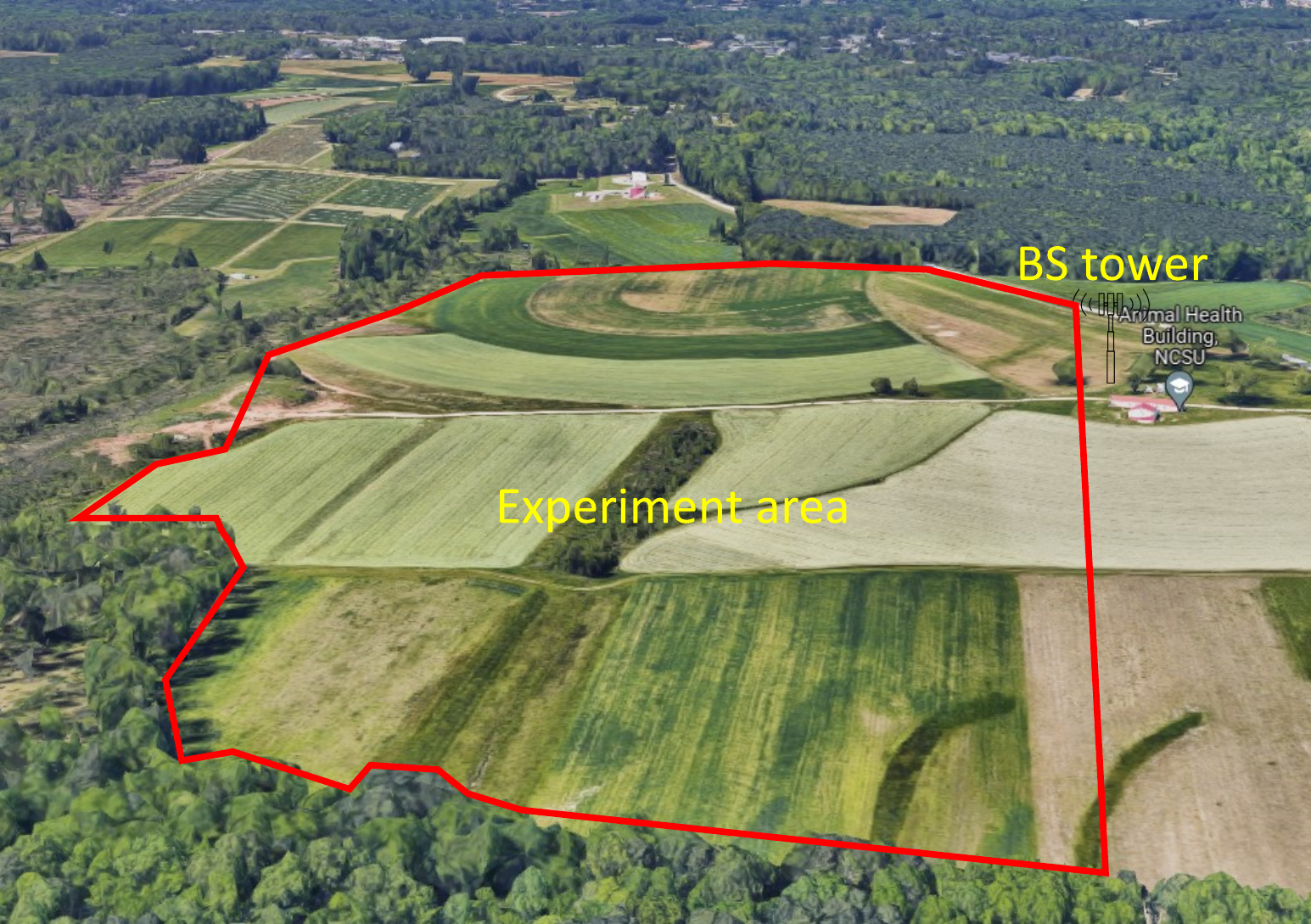}\label{fig:exp_site}}
 
        \subfloat[BS tower]{\includegraphics[width=0.175\textwidth]{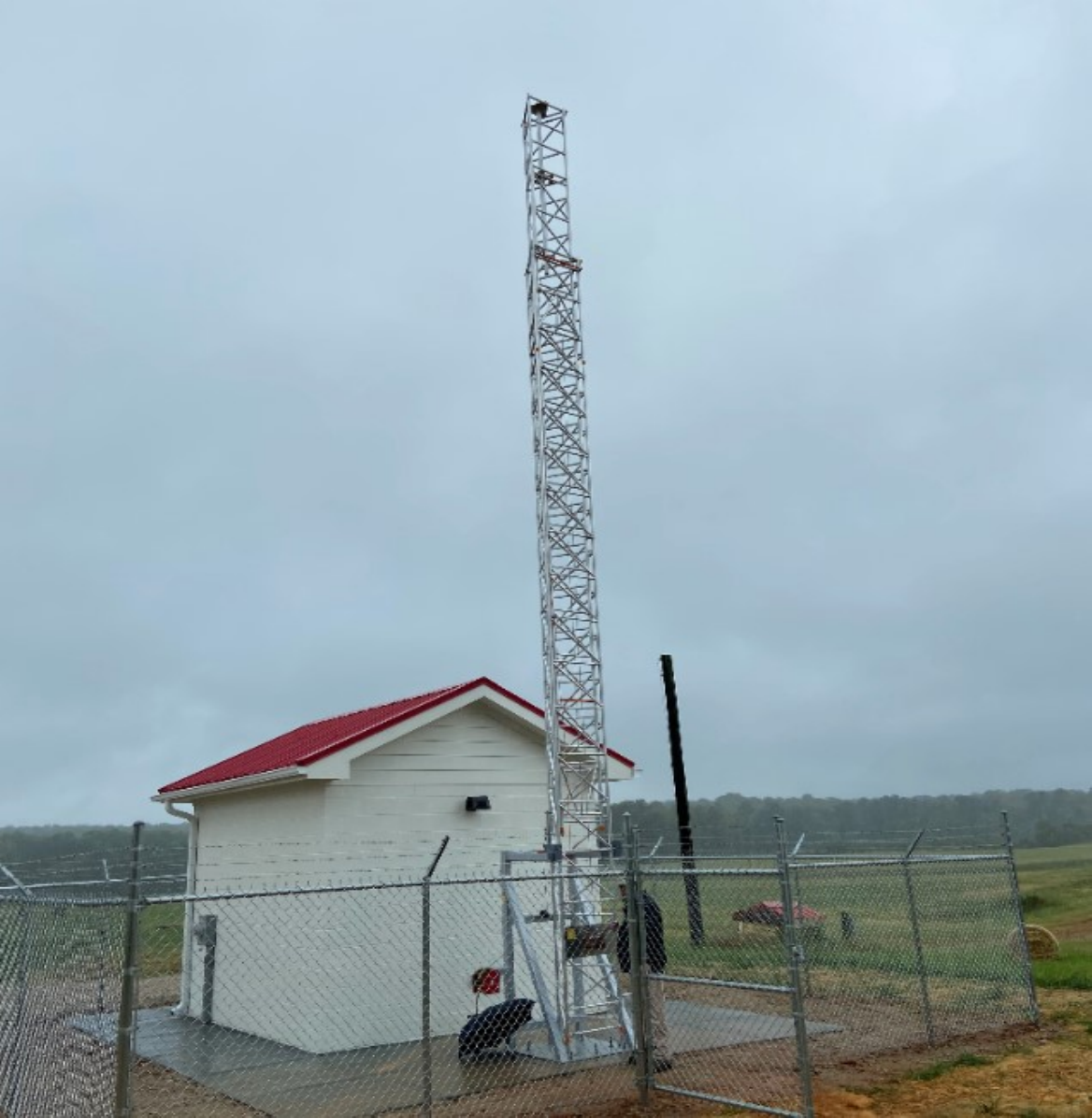}\label{fig:BS}}~
	\subfloat[Drone]{\includegraphics[width=0.30\textwidth]{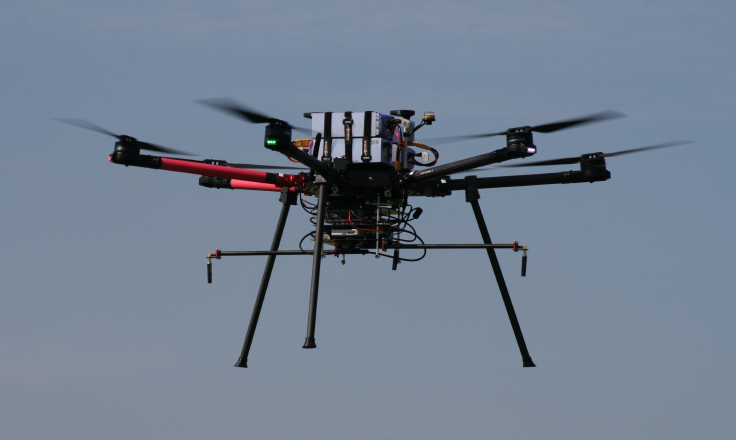}\label{fig:drone}}
	\caption{(a) The area where air-to-ground propagation data has been collected in AERPAW, (b) The fixed node tower (30 feet high) that includes the SDR serving as the LTE eNB, and (c) the drone that carries the receiver SDR.}
\end{figure}

\section{Measurement Campaign Overview}\label{sec:meas_camp}
In this section, we describe the details of our radio propagation measurements. We present our measurement setup, define UAV trajectory used, and describe our approach for characterizing antenna effects.

\subsection{Measurement Setup}
% \begin{figure}[t]
% 	\centering
% 	\subfloat[BS tower]{\includegraphics[width=0.2\textwidth]{BS_photo.pdf}\label{fig:BS}}~
% 	\subfloat[Drone]{\includegraphics[width=0.25\textwidth]{drone_photo.pdf}\label{fig:drone}}
% 	\caption{(a) The fixed node tower (30 feet high) that includes the SDR serving as the LTE eNB and (b) the drone that carries the receiver SDR.}\label{fig:BS_drone}
% \end{figure}
The measurement campaign was conducted at the Lake Wheeler Road Field Labs (LWRFL) site in Raleigh, NC, USA, which is one of the two sites in the NSF Aerial Experimentation and Research Platform for Advanced Wireless (AERPAW). The experimental area, depicted in Fig.~\ref{fig:exp_site}, can be classified as an open rural environment, ensuring LoS conditions between a UAV and the BS throughout the entire duration of the experiments. Fig.~\ref{fig:BS} and Fig.~\ref{fig:drone} present photos of the base station (BS) tower and the drone used during the measurement campaign. The BS tower stands at a height of 10 meters and is equipped with a single dipole transmit antenna. On the other hand, the drone is equipped with a vertically oriented single dipole receiver antenna and a GPS receiver to accurately track its position. To facilitate the measurements, the srsRAN open-source Software Defined Radio (SDR) software was utilized to implement an LTE evolved NodeB (eNB) at the BS tower, as shown in Fig.~\ref{fig:BS}. The eNB continuously transmitted common reference symbols (CRSs) during the measurement campaign. 

During the measurement campaign, the drone collects raw I/Q data samples using a Software Defined Radio (SDR) that is attached to it. Specifically, the USRP B205mini from National Instruments (NI) is utilized as the SDR device, both at the BS tower and on the UAV. For post-processing the raw I/Q data, we employ Matlab's LTE toolbox. Within this toolbox, we calculate the Reference Signal Received Power (RSRP) for each location of the UAV. To ensure efficient processing and analysis, we collect 20 ms segments of data out of every 100 ms. Within each 20 ms segment, we extract a 10 ms duration for subsequent post-processing. Throughout the paper, the terms ``received signal" and ``RSRP" are used interchangeably to refer to the measured signal strength. The major specifications of the transmitter and the receiver are listed in Table I.

\begin{table}[t]
\renewcommand{\arraystretch}{1.1}
\caption{Measurement Setup for Experiments}
\label{table:settings}
\centering
\begin{tabular}{lc}
\hline
BS Tower (Transmitter) \\
\hline
Technology & LTE\\
Tower height & 10~m\\
Transmit power & 10~dBm\\
Carrier frequency & 3.51~GHz\\
Bandwidth & 1.4~MHz\\
Antenna & Dipole antenna (RM-WB1)\\
\hline
\hline
UAV (Receiver) \\
\hline
Antenna & Dipole antenna (SA-1400-5900)\\
UAV heights & \{30,~50,~70,~90,~110\}~m\\
\hline
\end{tabular}
\vspace{-0.15in}
\end{table}

\subsection{UAV Trajectory}
\begin{figure}[t]
	\centering
	\subfloat[Top view of trajectory and RSRPs ($h=110$~m).]{\includegraphics[width=0.48\textwidth]{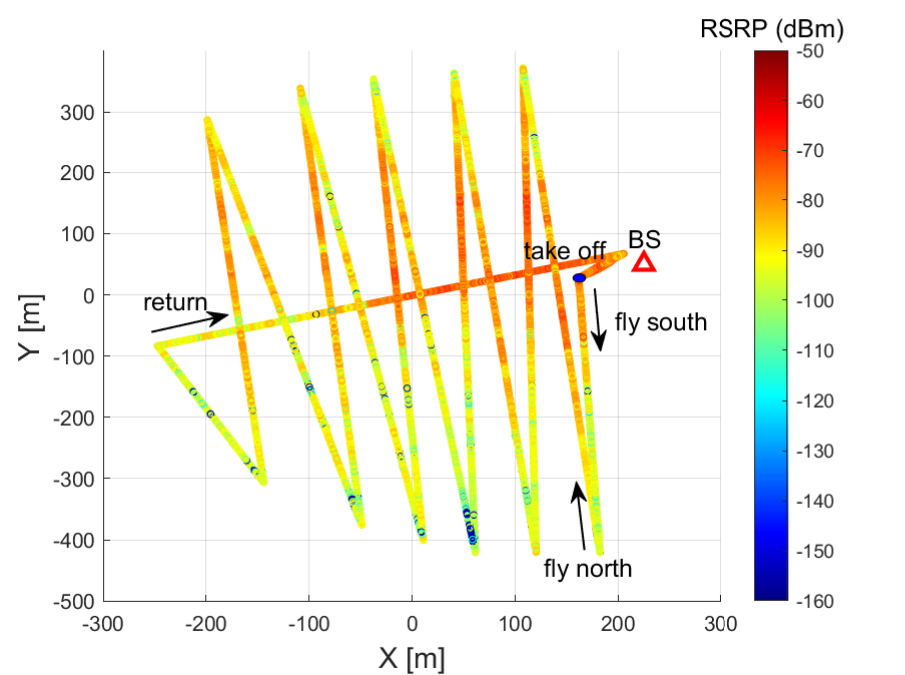}}
	
	\subfloat[3D view of trajectory and RSRP.]{\includegraphics[width=0.48\textwidth]{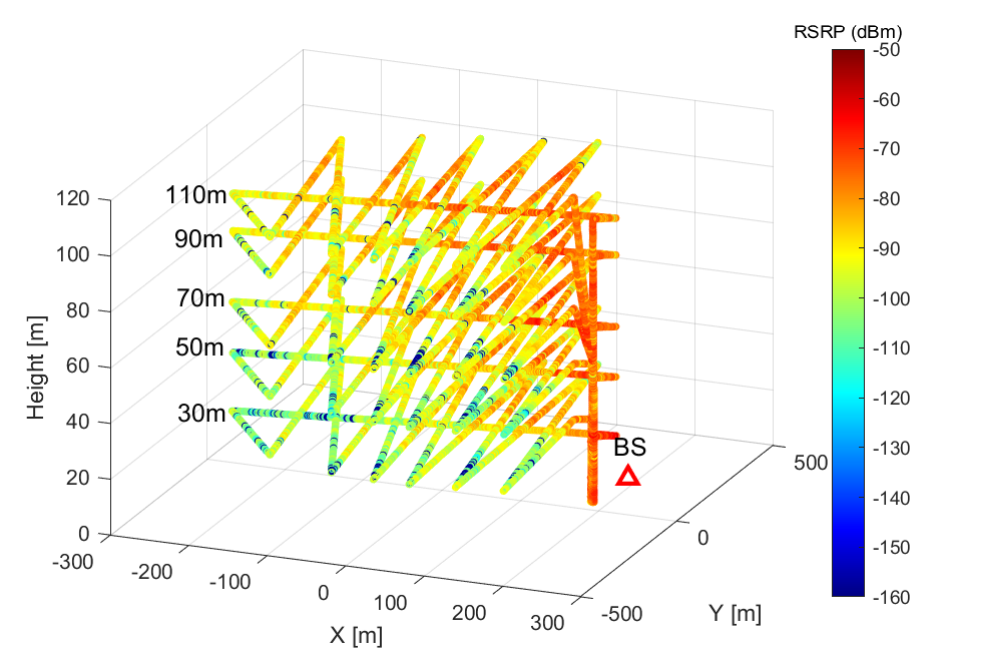}}
	\caption{The trajectory of the UAV and measured RSRP in RDZ experiments. Multiple flights with different heights are plotted in the bottom figure.}\label{fig:trajectory}
\end{figure}
We conduct the experiments multiple times by changing the altitude (height) of the UAV from 30~m to 110~m at increments of 20 m. In each flight, the UAV flies an identical predefined trajectory 
%\textcolor{red}{on the bottom at the plane but} 
with a different fixed height. In particular, the UAV flies on a zig-zag pattern through the experiment site, between south and north waypoints, and it eventually flies back to the starting point. The top view (at $h=110$~m) and the 3D view of the UAV trajectories along with measured RSRPs are illustrated in Fig.~\ref{fig:trajectory} for flight trajectories at $30$~m, $50$~m, $70$~m, $90$~m, and $110$~m.  %After the UAV takes off, it flies south and north in a zigzag pattern. Then, it returns to the starting point. 
%In the top view result, we can observe that the trajectories of the different heights are fully overlapped as they are programmed.

\subsection{Antenna Radiation Pattern Characterization}\label{sec:ant_pat}
\begin{figure}[t!]
	\centering
	\subfloat[Anechoic chamber setup for Tx antenna pattern measurement. The center point of the chamber is adjusted to the tip of the antenna by the crossed laser lines.]{\includegraphics[width=0.45\textwidth]{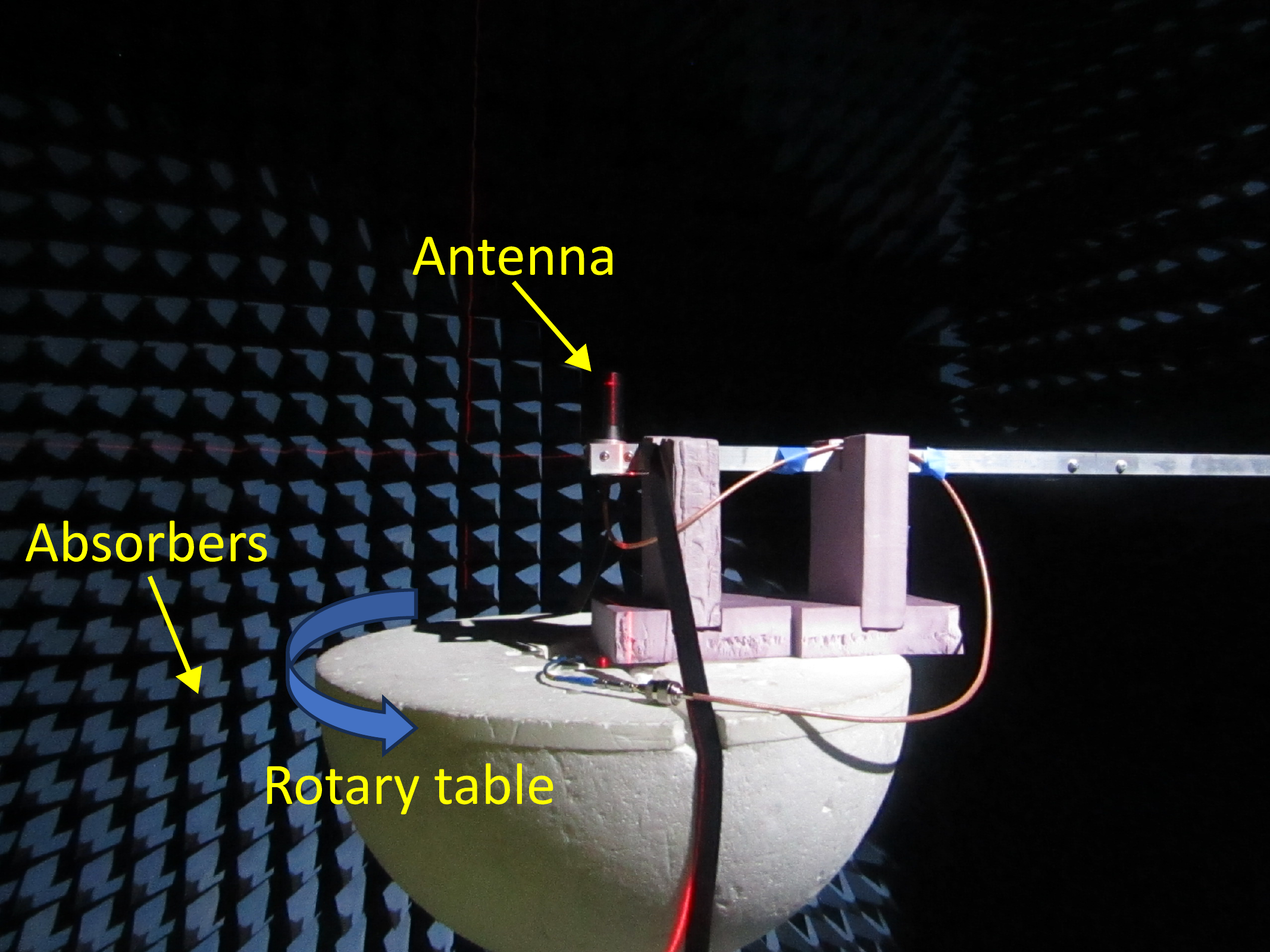}\label{fig:chamber}}
        
        \subfloat[Measured Tx antenna pattern in Cartesian coordinates at 3.5~GHz (by the gain in linear scale).]{\includegraphics[width=0.24\textwidth]{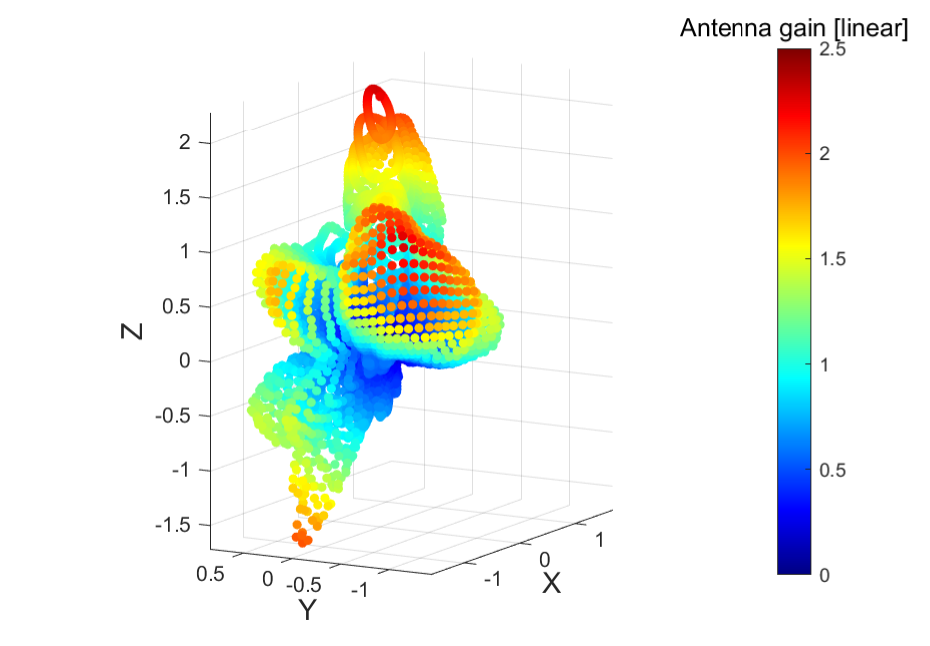}\label{fig:Tx_ant_pat}}~
        \subfloat[Rx antenna pattern for the elevation angle domain at 2.4~GHz (gain in dB scale from the specification sheet).]{\includegraphics[width=0.24\textwidth]{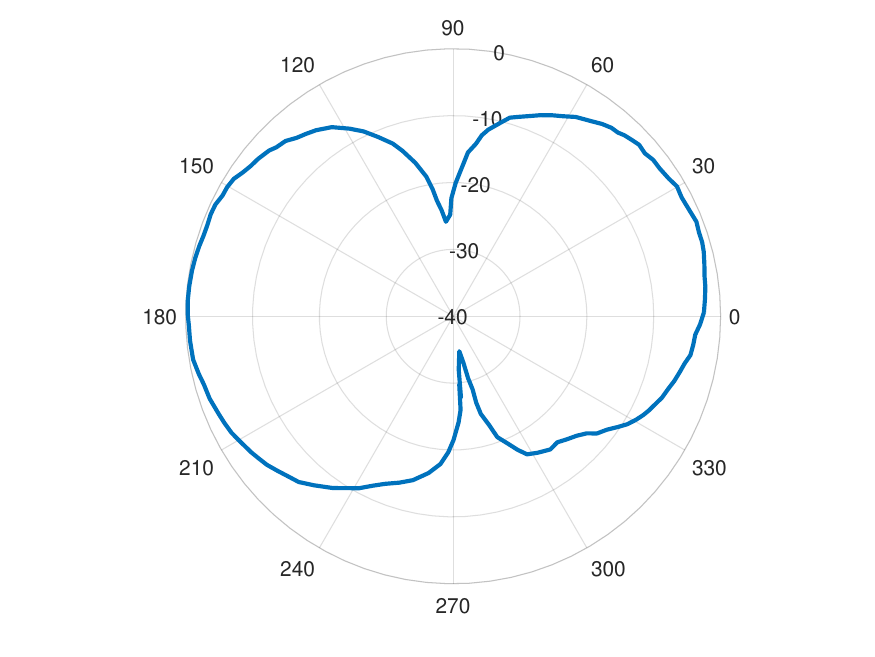}\label{fig:Rx_ant_pat}}
        
        \subfloat[Combined (Tx $+$ Rx) antenna pattern in 2D angle domain in dB scale. Only the angle space in the black rectangle is used for LoS between the BS tower and the UAV in the experiment ($\theta_{\rm l}$), and the red rectangle is used for the ground reflection angles ($\theta_{\rm r}$).]{\includegraphics[width=0.48\textwidth]{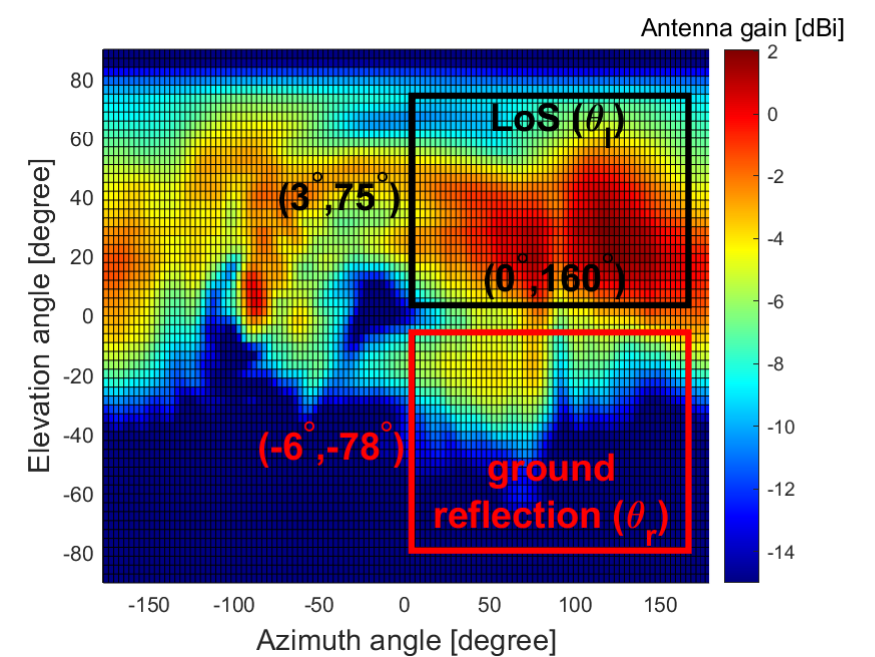}\label{fig:tot_ant_pat}}
	\caption{The anechoic chamber setup for the 3D antenna pattern measurement and the Tx, Rx, and combined antenna patterns we used for the analysis. }\label{fig:ant_pat}
\end{figure}
\begin{figure*}[t!]
	\centering
	\subfloat[RSRP vs. time (measured antenna pattern).]{\includegraphics[width=0.33\textwidth]{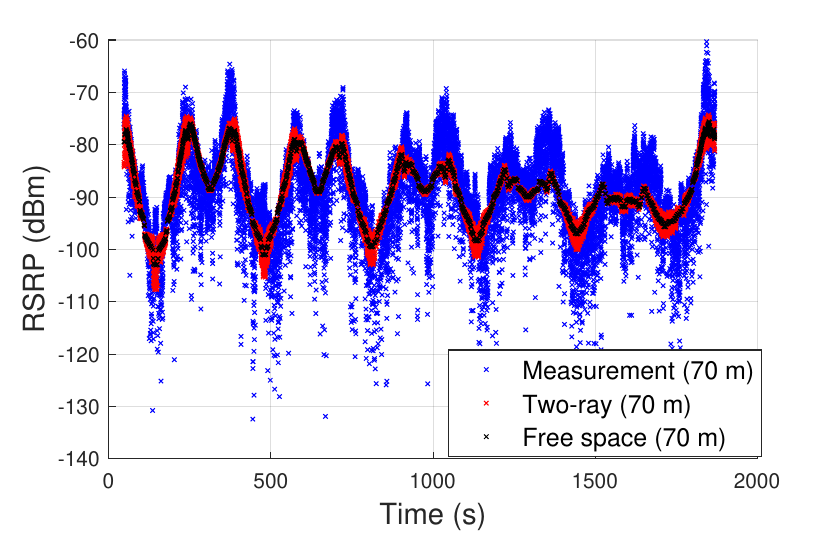}\label{fig:RSRP_fit_ant_t_m}}~
	\subfloat[RSRP vs. 3D distance (measured antenna pattern).]{\includegraphics[width=0.33\textwidth]{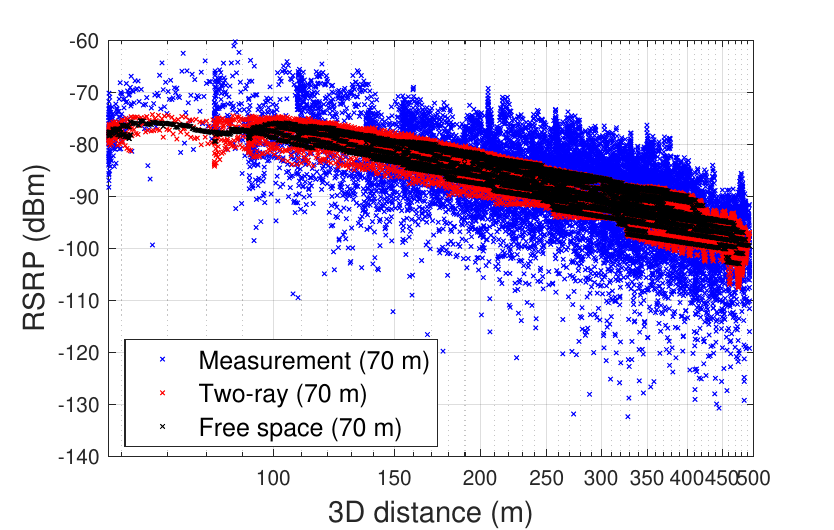}\label{fig:RSRP_fit_ant_d_m}}~
        \subfloat[RSRP vs. elevation angle (measured antenna pattern).]{\includegraphics[width=0.33\textwidth]{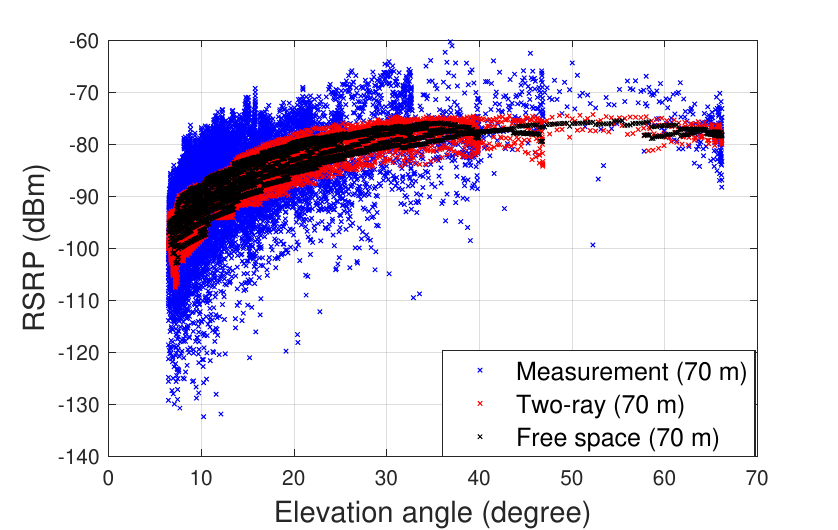}}

        \subfloat[RSRP vs. time (dipole antenna pattern).]{\includegraphics[width=0.33\textwidth]{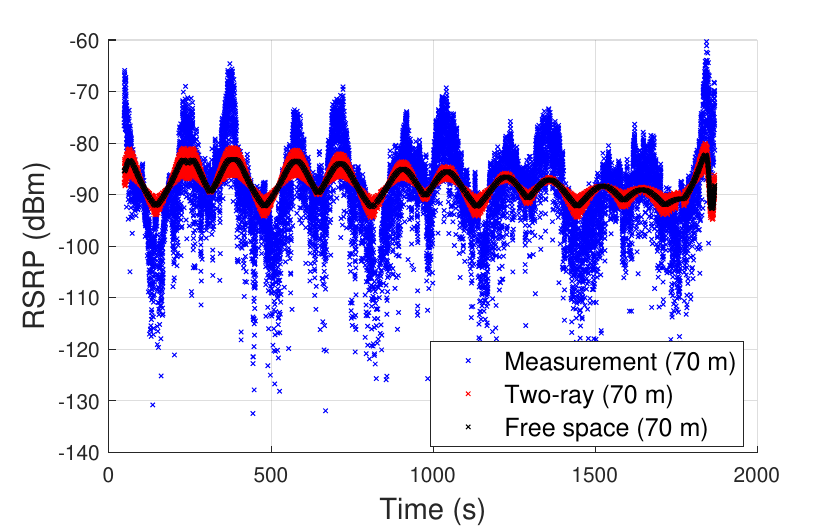}}~
	\subfloat[RSRP vs. 3D distance (dipole antenna pattern).]{\includegraphics[width=0.33\textwidth]{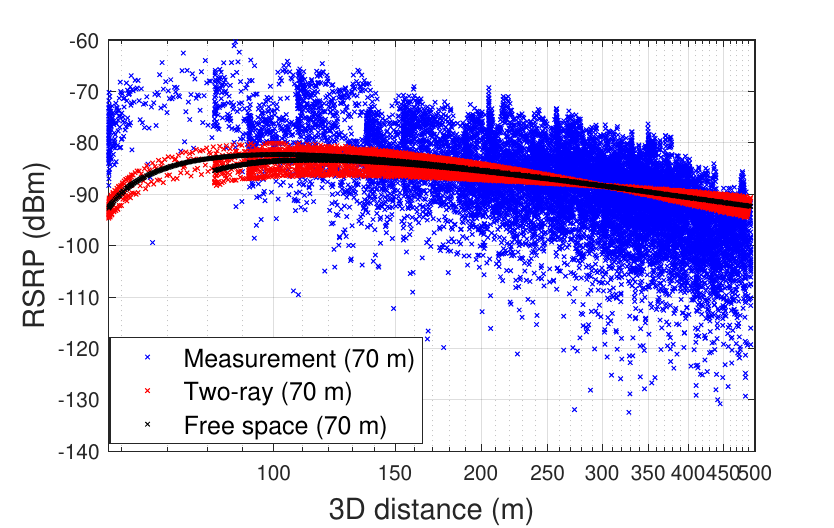}}~
        \subfloat[RSRP vs. elevation angle (dipole antenna pattern).]{\includegraphics[width=0.33\textwidth]{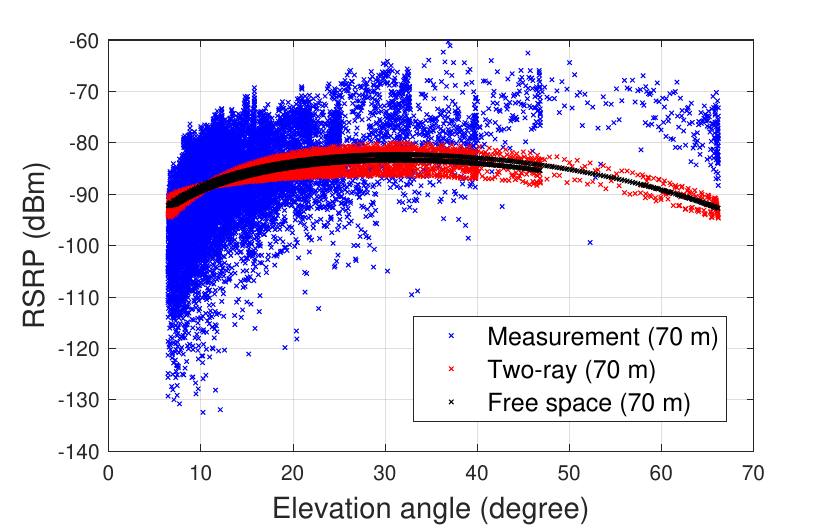}}

        \subfloat[RSRP vs. time (onmi-directional antenna pattern).]{\includegraphics[width=0.33\textwidth]{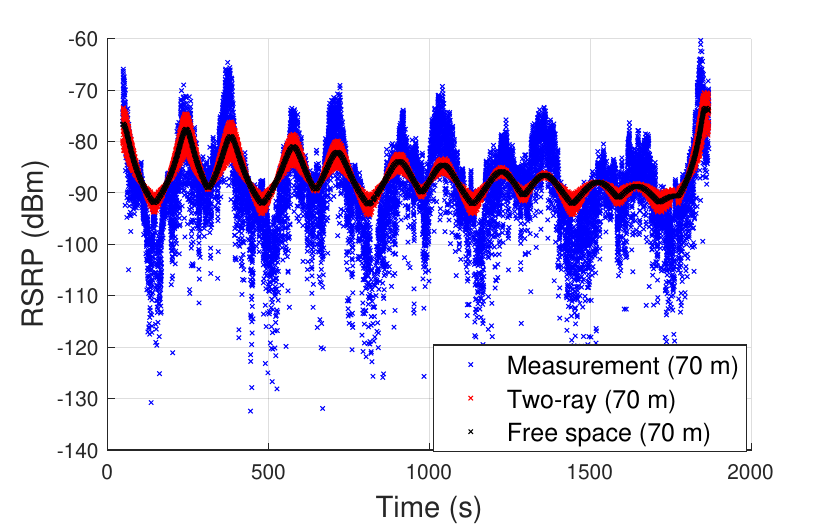}}~
	\subfloat[RSRP vs. 3D distance (onmi-directional antenna pattern).]{\includegraphics[width=0.33\textwidth]{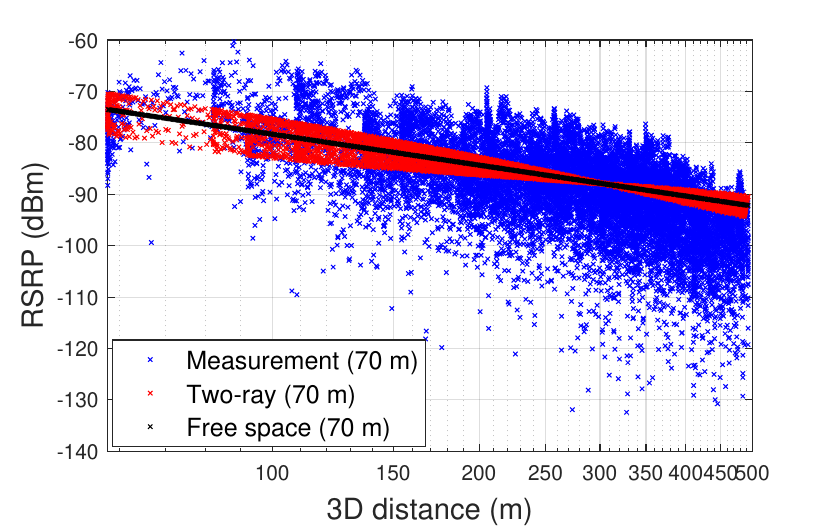}}~
        \subfloat[RSRP vs. elevation angle (onmi-directional antenna pattern).]{\includegraphics[width=0.33\textwidth]{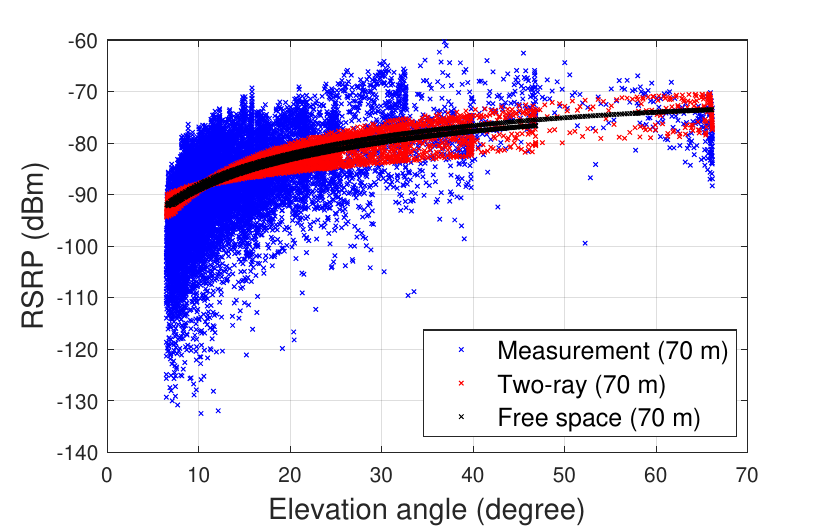}}
	\caption{RSRP fitting with different path loss models and antenna patterns in time, distance, and elevation angle domains.}\label{fig:RSRP_fit_ant}
\end{figure*}
The dipole antenna used in our experiments generally exhibits omni-directional radiation patterns in the azimuth angle domain, but oval-shaped radiation patterns in the elevation angle domain. The radiation pattern also varies with the carrier frequency. We obtained the antenna pattern specifications for the Rx dipole antenna (SA-1400-5900) from the vendor's specification sheet, and it shows a typical donut-shaped dipole pattern that remains consistent across different carrier frequencies~\cite{sa_1400}. Specifically, in the specification sheet, the antenna patterns for 1.4, 1.7, 2.4, 4.4, and 5.8~GHz frequencies are provided and all of them have similar dipole patterns. Therefore, we adopted the 2.4~GHz frequency antenna pattern from the specification sheet for our analysis. However, the Tx dipole antenna (RM-WB1-DN) exhibited different elevation angle domain patterns depending on the carrier frequency and had an asymmetric pattern that did not guarantee omni-directionality in the azimuth angle domain~\cite{rm_wb1}. Furthermore, the specification sheet did not provide the radiation pattern for the specific carrier frequency (3.51 GHz) used in our experiments. To obtain the exact antenna radiation pattern for the 3.51~GHz frequency, we conducted separate measurements of the 3D antenna pattern using an anechoic chamber facility located at wireless research center (WRC), Wake Forest, NC.

Fig.~\ref{fig:chamber} shows a photo of the setup in the anechoic chamber during the measurement of the Tx antenna's 3D pattern. Fig.~\ref{fig:Tx_ant_pat} displays the output of the antenna measurement, visualizing the antenna pattern in 3D Cartesian coordinates. It can be observed that the antenna pattern is not purely omni-directional in the azimuth angle domain, and the directivity in the elevation angle domain is not straightforward. In contrast, Fig.~\ref{fig:Rx_ant_pat} shows the elevation angle domain antenna pattern of the Rx antenna as provided in the specification sheet, where the antenna pattern is specified as omni-directional with uniform gain in the azimuth domain. Fig.~\ref{fig:tot_ant_pat} illustrates the combined antenna gain from the Tx and Rx antenna patterns from Fig.~\ref{fig:Tx_ant_pat} and Fig.~\ref{fig:Rx_ant_pat}, respectively, represented in the azimuth and elevation angle domain. For all UAV heights in our experiments, the LoS angles between the Tx tower and the UAV were within the angle space covered by the black rectangular area, while the ground reflection angles between Tx tower and the UAV were covered by the red rectangular area, which are illustrated in Fig.~\ref{fig:two_ray_illu}. This implies that the antenna pattern used for the analysis is limited to the angles within this space.

\section{Air-to-ground Propagation Modeling and Analysis}\label{sec:A2G_ana}
In this section, we review how we post-process the data for correcting errors in altitude reported by the UAV's GPS. Subsequently, we model the measured RSRP using different 3D propagation models that take into account two-ray multipath model and 3D antenna pattern.

\begin{figure}[t!]
	\centering
	\subfloat[CDF of RSRP.]{\includegraphics[width=0.48\textwidth]{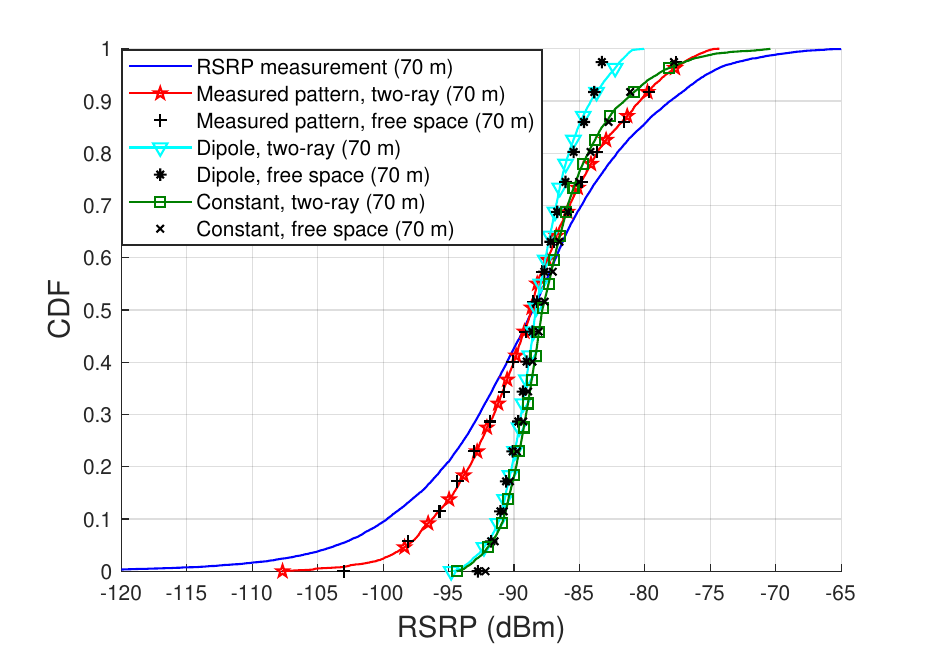}\label{fig:cdf_fit_RSRP}}
	\vspace{-0.0in}
	\subfloat[CDF of fitting error between measured RSRP and path loss models.]{\includegraphics[width=0.48\textwidth]{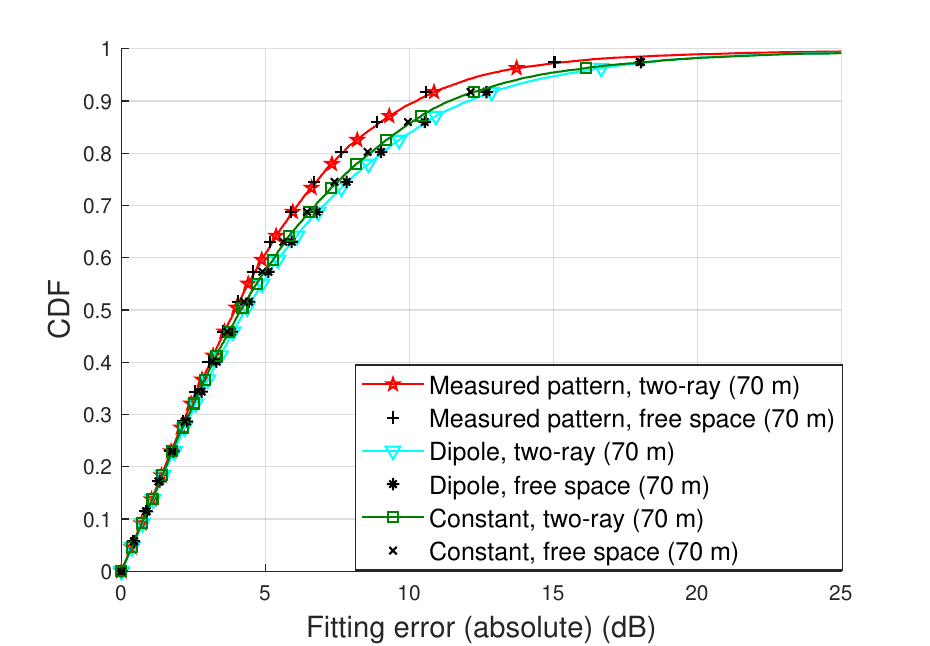}\label{fig:cdf_fit_err}}
	\caption{CDF of RSRP and the fitting error with path loss models by different antenna radiation patterns. The measured antenna pattern in Fig.~\ref{fig:ant_pat} achieves the closest fitting to the RSRP measurements.}\label{fig:cdf_fit}
\end{figure}
\begin{figure}[t!]
	\centering
	\subfloat[Fitting error vs. time]{\includegraphics[width=0.48\textwidth]{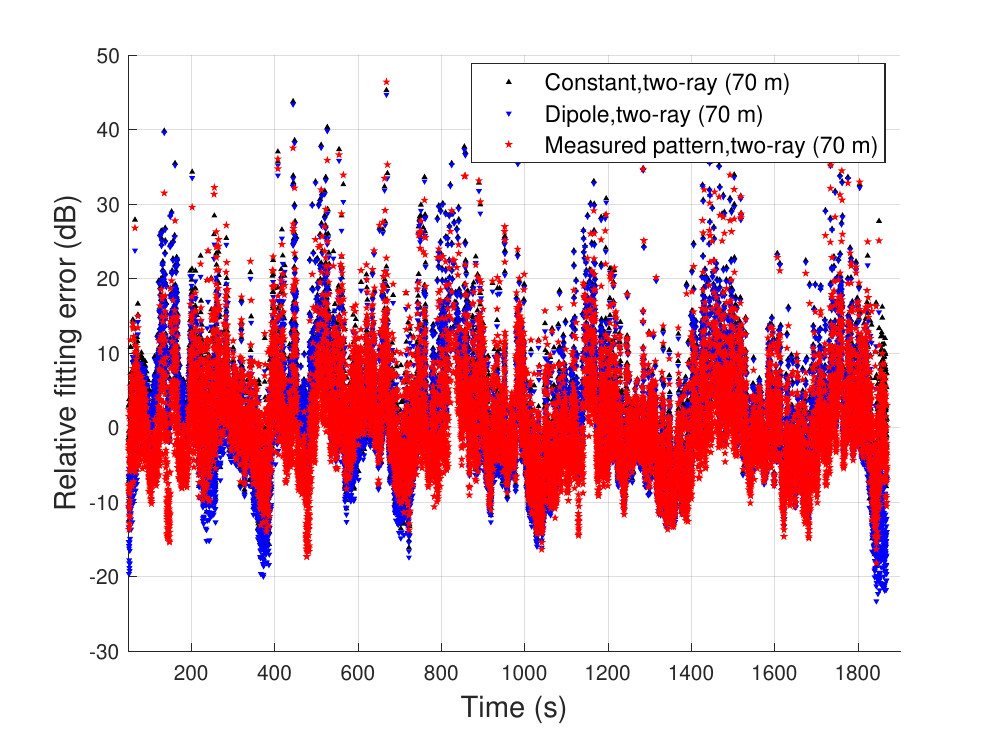}\label{fig:err_t_70}}
	\vspace{-0.0in}
	\subfloat[Fitting error vs. 3D distance]{\includegraphics[width=0.48\textwidth]{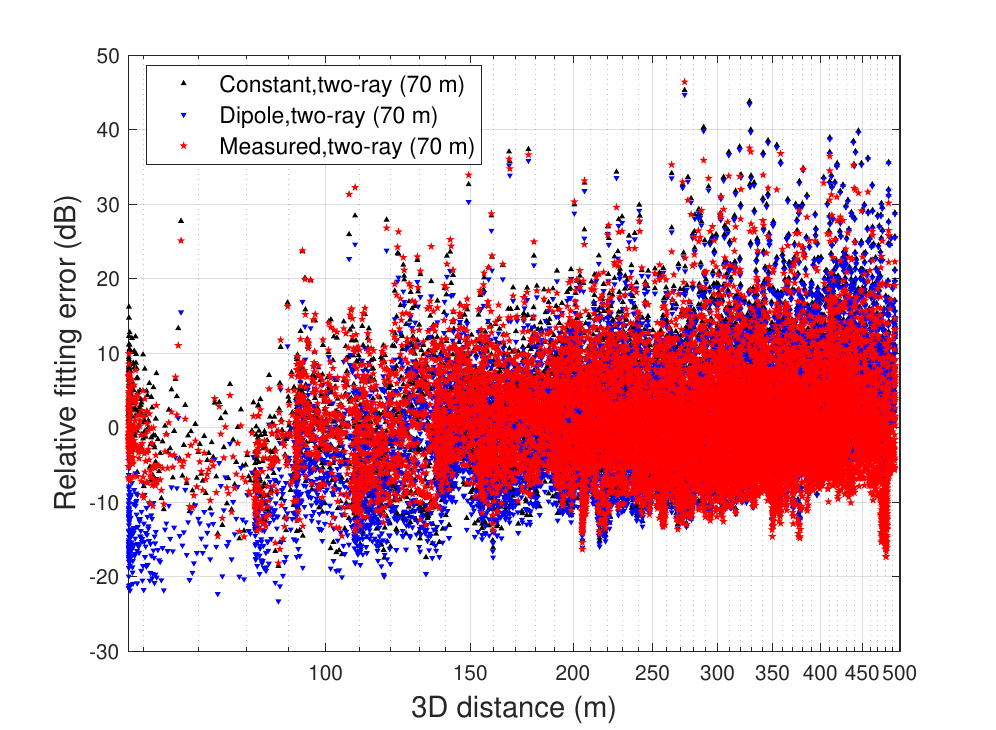}\label{fig:err_d_70}}
 	\vspace{-0.0in}
	\subfloat[Fitting error vs. elevation angle]{\includegraphics[width=0.48\textwidth]{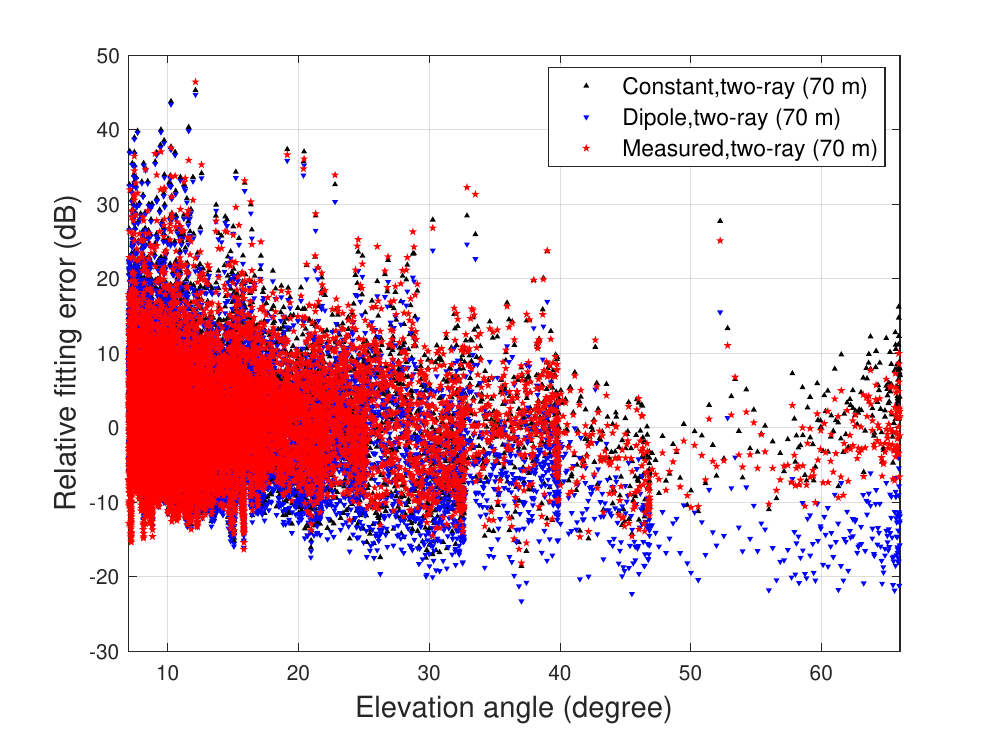}\label{fig:err_e_70}}
	\caption{Relative fitting error between measured RSRP and the two-ray path loss model with different antenna patterns in time, distance, and elevation angle domains.}\label{fig:err_70}
\end{figure}

\subsection{Post-measurement Correction of Altitude and RSRP}

During the measurements, we encountered calibration errors caused by limitations in the SDR hardware. Specifically, the Universal Software Radio Peripheral (USRP) mounted on the UAV exhibited a power level calibration error, resulting in a constant offset power throughout the experiment. To address this issue, we conducted a separate experiment to measure and determine the offset at the USRP, which was found to be 98~dB. Subsequently, we added this offset to the calculated RSRP values obtained from subsequent experiments, effectively compensating for the calibration offset.

Additionally, the GPS receiver carried by the UAV exhibited an altitude mismatch. We observed an altitude drift of approximately 6 m after the UAV landed, when compared with the initial altitude of the UAV. To rectify this mismatch, we applied a linear compensation approach (see~\cite[Fig.~6]{maeng2023aeriq}). This involved adjusting the altitude measurements such that the altitude at the end of the flight matched the altitude of the initial measurement. By applying this compensation, we aimed to ensure accurate altitude data throughout the experiment.

\begin{figure}[t!]
	\centering
	\subfloat{\includegraphics[width=0.38\textwidth]{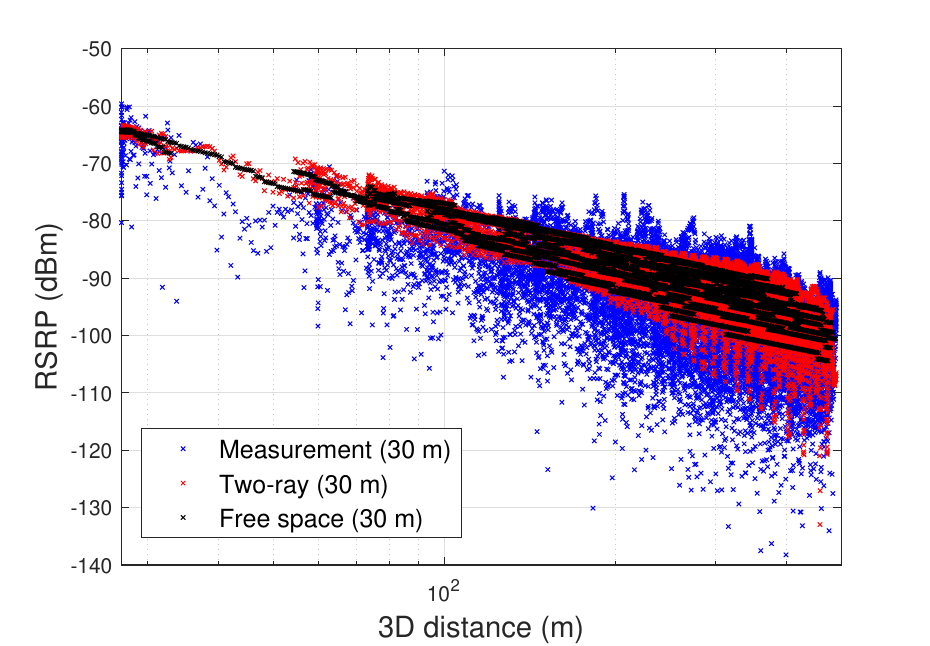}}
 
	\subfloat{\includegraphics[width=0.38\textwidth]{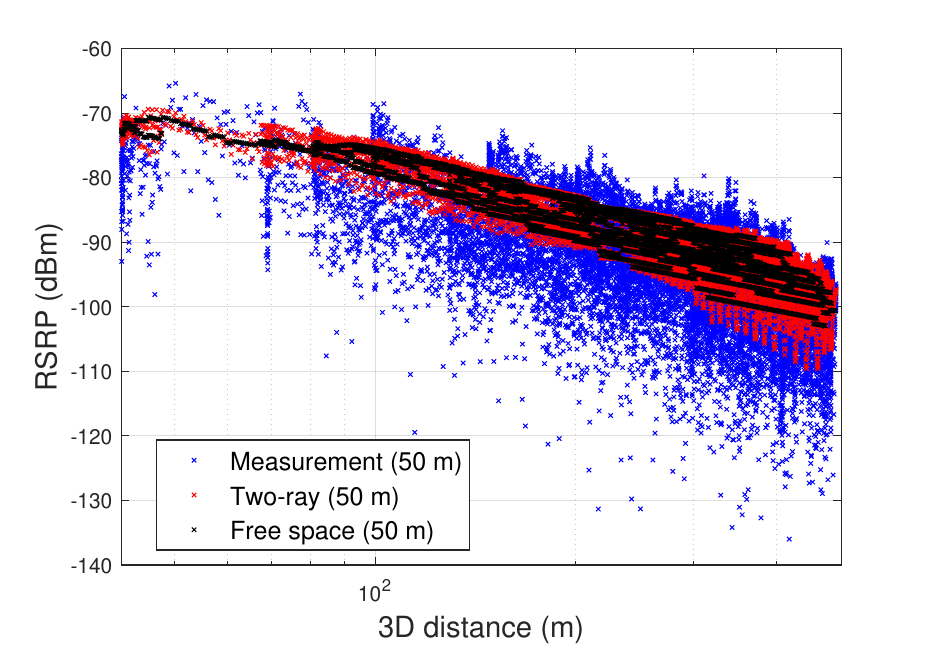}}

	\subfloat{\includegraphics[width=0.38\textwidth]{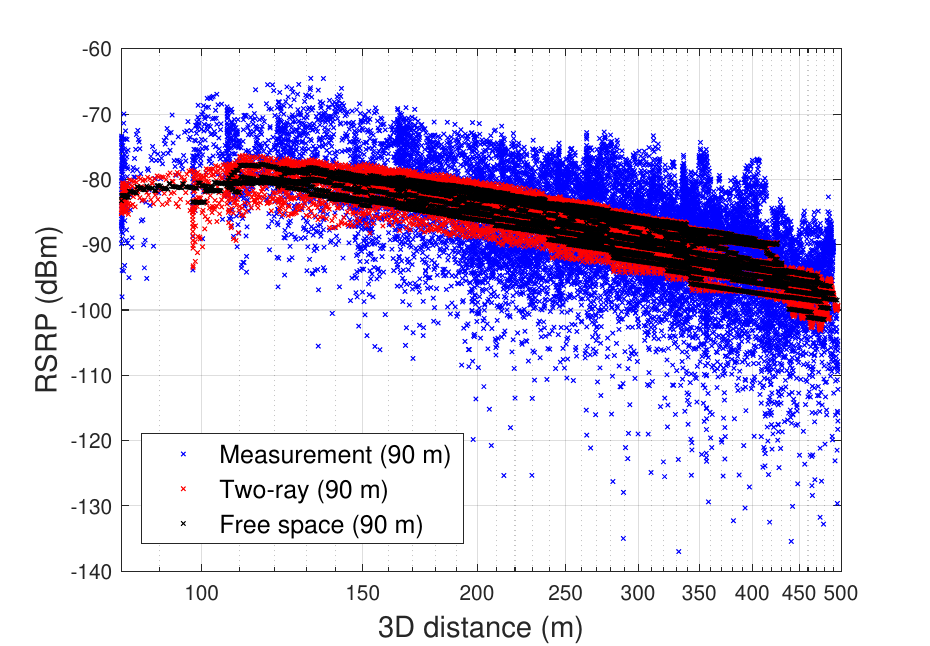}}
	
	\subfloat{\includegraphics[width=0.38\textwidth]{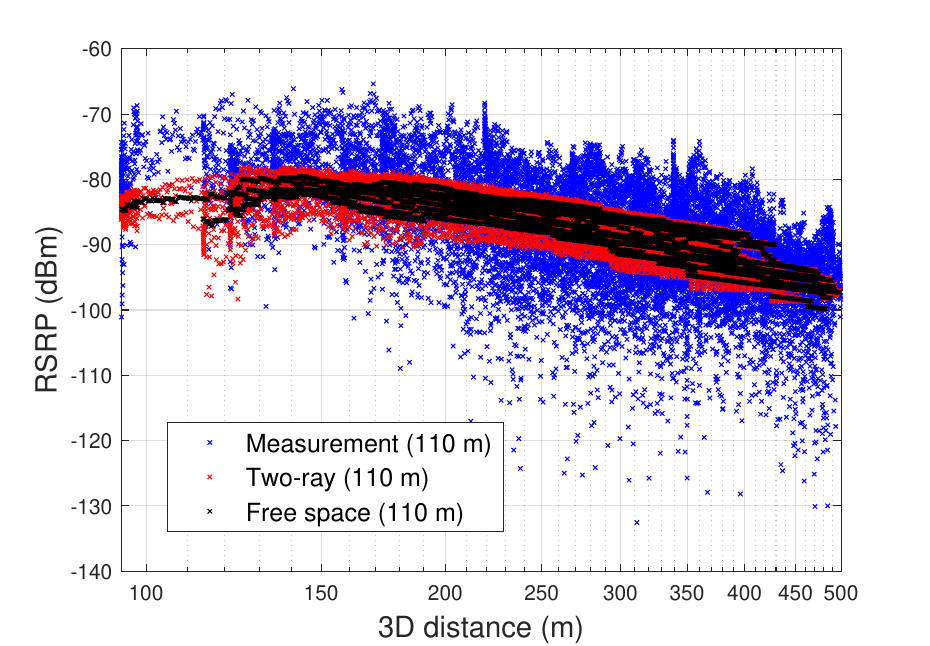}}
	\caption{Curve fitting for the RSRP measurements at different altitudes, considering free-space path loss and two-ray (ground-reflection) model.}\label{fig:PL_fit}
\end{figure}
\begin{figure}[t!]
	\centering
	\subfloat[Fitting error vs. 3D distance (measured antenna pattern)]{\includegraphics[width=0.48\textwidth]{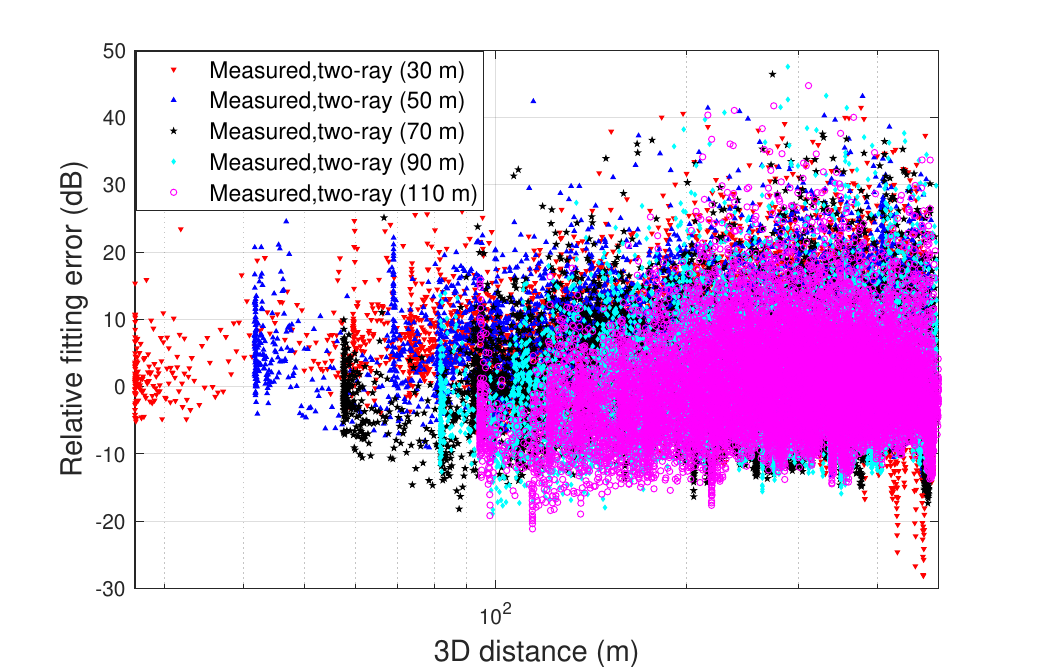}\label{fig:err_d_mea}}
	\vspace{-0.0in}
	\subfloat[Fitting error vs. 3D distance (dipole antenna pattern)]{\includegraphics[width=0.48\textwidth]{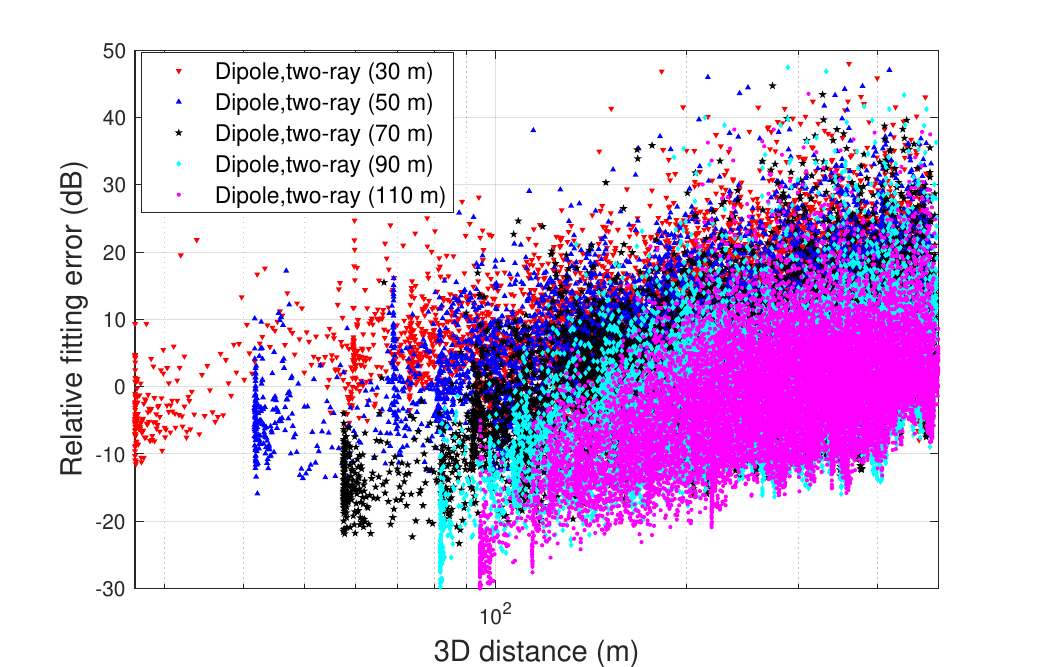}\label{fig:err_d_dip}}
 	\vspace{-0.0in}
	\subfloat[Fitting error vs. 3D distance (constant antenna pattern)]{\includegraphics[width=0.48\textwidth]{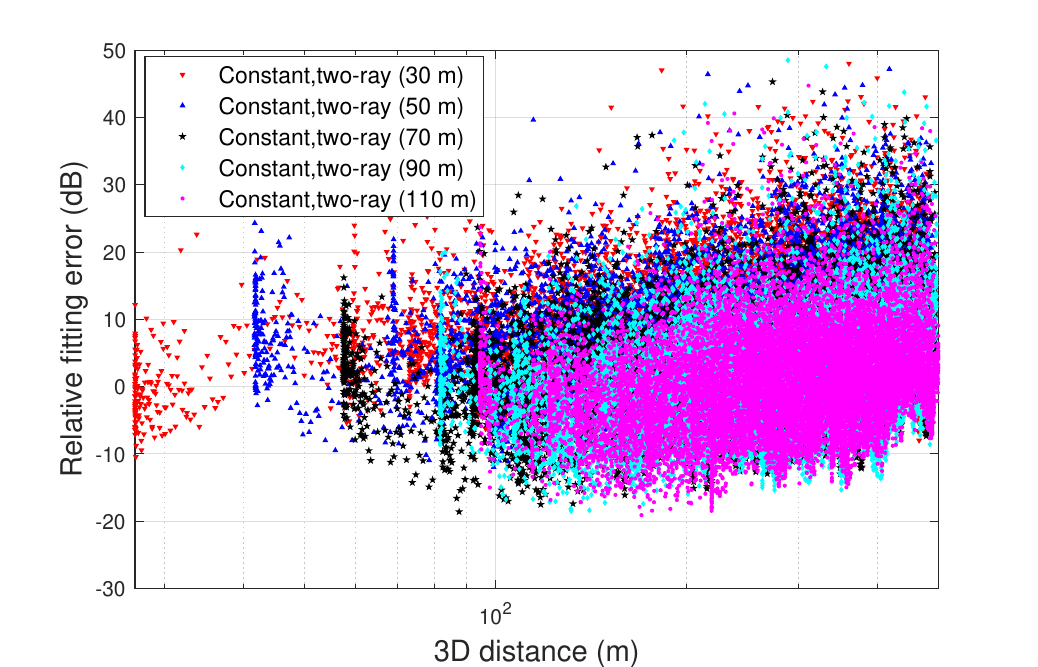}\label{fig:err_d_con}}
	\caption{Relative fitting error between measured RSRP and the two-ray path loss model with different antenna patterns and heights in distance domains.}\label{fig:err_d_all}
\end{figure}

\subsection{Antenna Radiation Pattern Effect in Path Loss Analysis}\label{sec:ant_ana}
In this subsection, we analyze the effect of antenna radiation patterns on the path loss fitting to the RSRP from the experiments. We consider three different antenna pattern setups for comparison: 1) Tx and Rx 3D antenna patterns described in Section~\ref{sec:ant_pat} and Fig.~\ref{fig:ant_pat}; 2) the donut shape dipole antenna pattern using the formulation for both Tx and Rx antennas; and 3) constant azimuth and elevation antenna gain for both Tx and Rx antennas. The dipole antenna pattern formula in the second case is given by~\cite{maeng2020interference}
\begin{align}\label{eq:dipole_formula}
\mathsf{G}_{\rm bs}(\theta)&=\mathsf{G}_{\rm uav}(\theta)=\frac{\cos\left(\frac{\pi}{2}\cos\theta\right)}{\sin\theta}.
\end{align}

Fig.~\ref{fig:RSRP_fit_ant} and Fig.~\ref{fig:cdf_fit} provide a comprehensive analysis of the RSRP fitting results using different antenna patterns and path loss models in \eqref{eq:PL_two}, \eqref{eq:PL_fs}. In Fig.~\ref{fig:RSRP_fit_ant}, the RSRP curves for a UAV height of 70 m are presented, along with the fitting results obtained from the free space and two-ray path loss models with different antenna patterns. It is observed that the antenna pattern described in Section~\ref{sec:ant_pat} provides the best fit to the RSRP curves, while the dipole pattern in \eqref{eq:dipole_formula} results in the worst fit. Additionally, Fig~\ref{fig:RSRP_fit_ant_t_m} highlights that the two-ray path loss model performs better than the free space path loss model in capturing the deep fading of RSRP.

To further evaluate the performance, Fig.~\ref{fig:cdf_fit} presents the cumulative distribution function (CDF) of the RSRP for the 70 m height measurement, along with the fitting results obtained from the path loss models and different antenna patterns. The CDF of the two-ray path loss model with the antenna pattern in Section~\ref{sec:ant_pat} matches closest with the CDF of the measured RSRP, indicating a better fit. Fig.~\ref{fig:cdf_fit_err} shows the fitting error, which is calculated by subtracting the measured RSRP from the fitted RSRP using the path loss models. It is observed that the fitting error is the smallest when using the two-ray path loss model with the antenna pattern in Section~\ref{sec:ant_pat}.

Fig.~\ref{fig:err_70} also evaluates the fitting error with different antenna patterns in time, distance, and elevation domains. It is observed that the dipole antenna pattern has the largest fitting error in short and long distances. We also observe that the fitting error is relatively high in small elevation angles. It implies that the effect of scattering from the objects around the test site increases the variance of the error when the elevation angle is low.
Overall, these results demonstrate that the choice of antenna pattern and path loss model significantly impacts the accuracy of RSRP fitting for air-to-ground communication links. The 3D antenna radiation pattern described in Section~\ref{sec:ant_pat}, combined with the two-ray path loss model, provides the best fit to the measured RSRP and minimizes the fitting error.

\subsection{Path Loss Model Fitting with Measurement}
Fig.~\ref{fig:PL_fit} illustrates the measured and fitted RSRP values as a function of 3D distance for different UAV heights ranging from 30 m to 110 m. We adopt measured antenna patterns in Section~\ref{sec:ant_pat}. The fitted curves follow the measured RSRP values reasonably closely. It is worth noting that the two-ray path loss model performs better in capturing the fluctuation of signal strength due to the ground reflected path compared to the free-space path loss model, especially when the UAV height is low. In the logarithmic scale of the distance domain, the RSRP is expected to decrease linearly. However, in the short distance range, a concave curve can be observed. This phenomenon is a result of the elevation-dependent antenna gain and the dramatic change in the elevation angle at short distances and high UAV altitudes. The 3D antenna pattern considered in the path loss models effectively captures this effect, leading to more accurate RSRP fitting. Overall, the results in Fig.~\ref{fig:PL_fit} highlight the importance of considering the elevation-dependent antenna gain and the 3D antenna pattern in accurately modeling and fitting RSRP measurements in air-to-ground communications.

Fig.~\ref{fig:err_d_all} shows the relative fitting error in the distance domain for all heights with different antenna patterns. The error by the dipole antenna pattern is relatively higher than other antenna patterns, especially when the distance is around 100~m to 200~m due to the antenna pattern mismatch. We also observe that the fitting error for the omnidirectional antenna pattern is higher than the measured antenna pattern for a large distance. Overall, the use of the measured antenna pattern results in the best fit for the measured data.

\subsection{Analysis of Shadowing Components from Measurement}\label{sec:anal_shad}
\begin{figure}[t]
	\centering
	\subfloat[30~m height]{\includegraphics[width=0.25\textwidth]{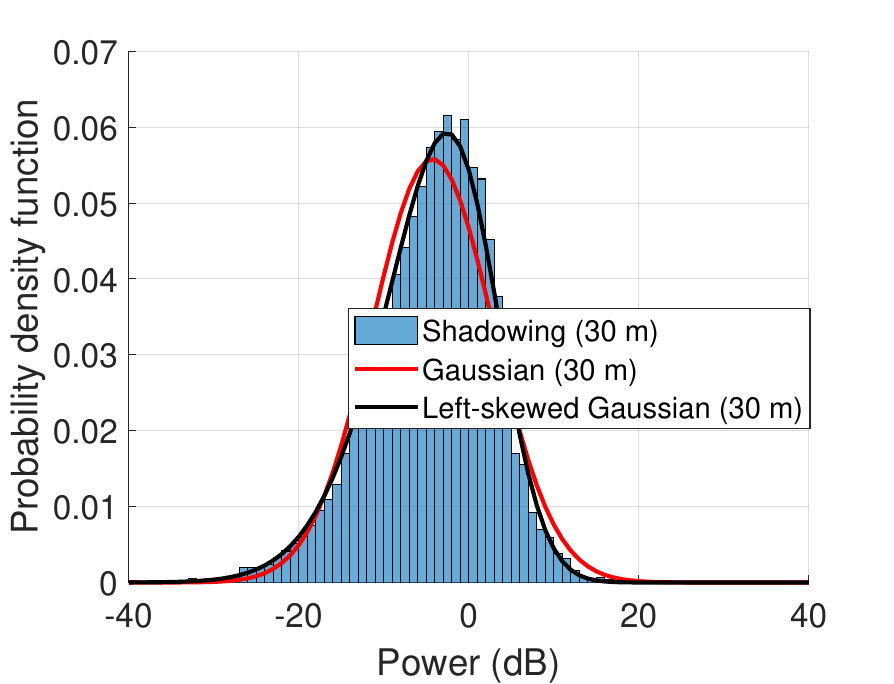}}~
	\subfloat[90~m height]{\includegraphics[width=0.25\textwidth]{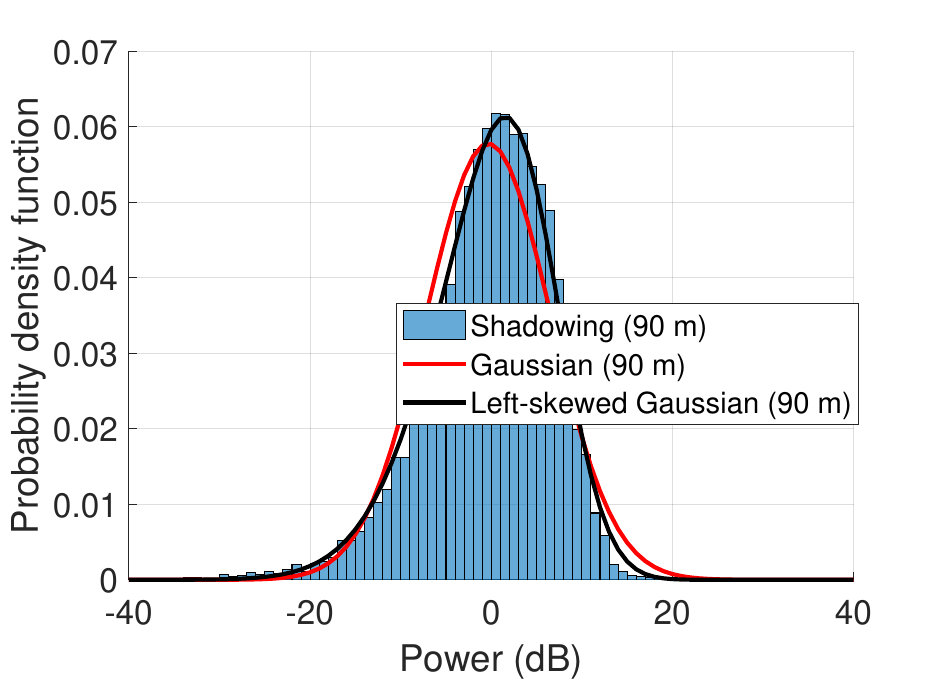}}
	\caption{Shadowing component from measurements and fitted curves to Gaussian and skewed Gaussian distribution, at (a) 30~m, and (b) 90~m UAV altitude.}\label{fig:shad_fit}
\end{figure}
After we derive the two-ray path loss model, we can extract the shadowing component by subtracting the path loss model from measured RSRP using \eqref{eq:received},  
% The distribution of the extracted shadowing component with different UAV heights is shown in Fig.~\ref{fig:shad_fit}. It is well-known that the shadowing is modeled by the Gaussian distribution, and we compare the measured shadowing with the fitted curve. We found that the distributions of the shadowing from the experiments are fit well to the Gaussian distribution, but those are not perfectly matched. Unlike the Gaussian distribution, which is symmetric, the measured distribution is asymmetric with a heavier left tail. Alternatively, we can fit the data better using a skewed Gaussian (normal) distribution, which can introduce a desired level of skewness to the Gaussian distribution~\cite{sung2017skew}. The probability density function (PDF) of the skewed Gaussian distribution can be written as
as shown in Fig.~\ref{fig:shad_fit}.
%shows the distribution of the extracted shadowing component obtained by subtracting the path loss model from the measured RSRP using \eqref{eq:received}. 
The shadowing component is known to follow a Gaussian distribution, and the measured shadowing distributions for different UAV heights are compared to the fitted curves. It is observed that the measured shadowing distributions can be modeled using a Gaussian distribution, though there are slight deviations. In particular, the measured distributions exhibit asymmetry with a heavier left tail compared to the symmetric Gaussian distribution. To achieve a better fit, an alternative approach is to use a skewed Gaussian (normal) distribution, which allows for introducing a desired level of skewness to the distribution~\cite{sung2017skew}. The probability density function (PDF) of the skewed Gaussian distribution can be expressed as
\begin{align}\label{eq:skew_Gau}
    f(x)=2\phi\left(\frac{x-\xi}{\omega}\right)\Phi\left(\alpha\left(\frac{x-\xi}{\omega}\right)\right),
\end{align}
where $\phi(\cdot)$, $\Phi(\cdot)$ indicates the PDF and the CDF of Gaussian distribution, respectively. The parameter $\alpha$ in \eqref{eq:skew_Gau} decides the skewness of the distribution. If $\alpha$ is a positive real value, it gives right-skewness, while left-skewness is introduced by a negative real value. In addition, the mean, the standard deviation of the shadowing, left-skewed Gaussian parameter $\alpha$, and normalized mean squared error (NMSE) of model fittings for all heights are listed in Table~II. Note that the optimal $\alpha$ is decided by minimizing NMSE. It shows that the distributions as well as the value of variances in different heights are similar, and we can assume a stationary process in spatial data.

\begin{table}[t]
\renewcommand{\arraystretch}{1.1}
\caption{Mean and standard deviation of the shadowing component for both measurement and Gaussian fitting curve, and NMSE of models fitting.}
\label{table:shad_stat}
\centering
\begin{tabular}{c | c | c | c | c | c}
\hline
\makecell{UAV \\ height} & Mean & \makecell{Standard\\ deviation} & $\alpha$ & \makecell{NMSE \\ (Gaussian)} & \makecell{NMSE \\ (skewed)} \\
\hline
$30$ m & $-4.26$~dB & $7.14$~dB & -2.13 & 0.0314 & 0.0027 \\
$50$ m & $-4.57$~dB & $6.45$~dB & -2.26 & 0.0370 & 0.0020  \\
$70$ m & $-0.34$~dB & $6.53$~dB & -2.57 & 0.0437 & 0.0036 \\
$90$ m & $-0.32$~dB & $6.90$~dB & -2.08 & 0.0338 & 0.0028 \\
$110$ m & $-0.27$~dB & $6.83$~dB & -2.27 & 0.0302 & 0.0022 \\
\hline
\hline
\end{tabular}
\vspace{-0.15in}
\end{table}

\section{Numerical Results on 3D Signal Interpolation}\label{sec:sim_Kriging}

In this section, we will first study the horizontal, vertical, and finally 3D correlation in the measured data. We will use the 3D correlation to calculate the semi-variogram, which will subsequently be used to analyze the 3D interpolation accuracy for various scenarios.  

\subsection{Analysis of Correlation Function from Measurement}
\subsubsection{Horizontal distance correlation}
In this subsection, we analyze the spatial correlation using the AERPAW datasets available at~\cite{IEEEDataPort}. We obtain correlation functions between two different 3D locations by using measurements at different heights, and we use exponential and bi-exponential functions to model the correlations as discussed earlier. The mean and standard deviation values obtained by statistical analysis in Section~\ref{sec:anal_shad} and the measured RSRP values are utilized in calculating the correlations.
\begin{figure}[t]
	\centering
	\subfloat[Correlation of horizontal distance.]{\includegraphics[width=0.25\textwidth]{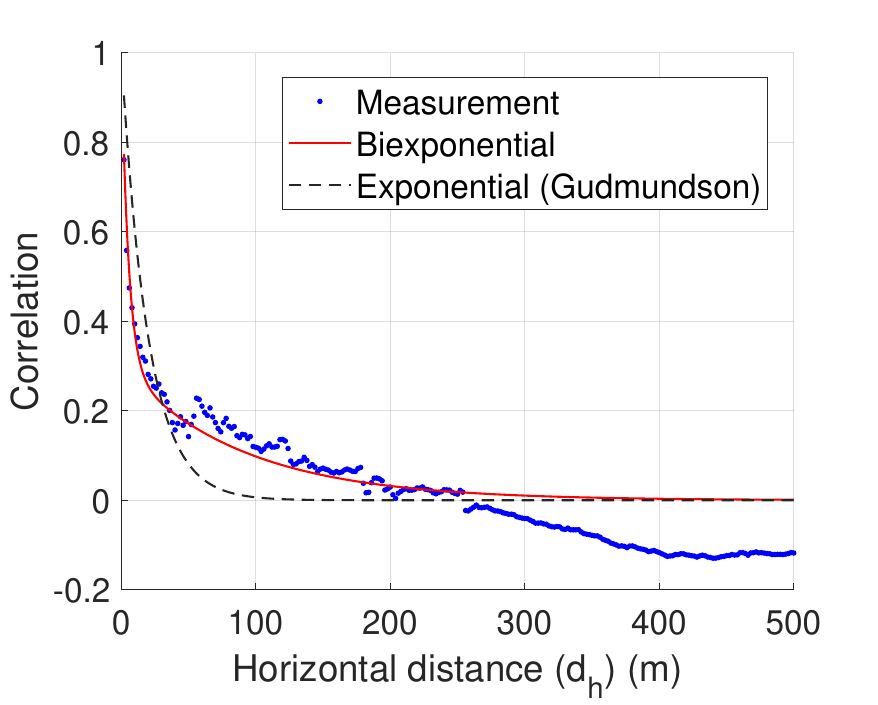}\label{fig:corr_dh}}~
        \subfloat[Correlation of vertical distance.]{\includegraphics[width=0.25\textwidth]{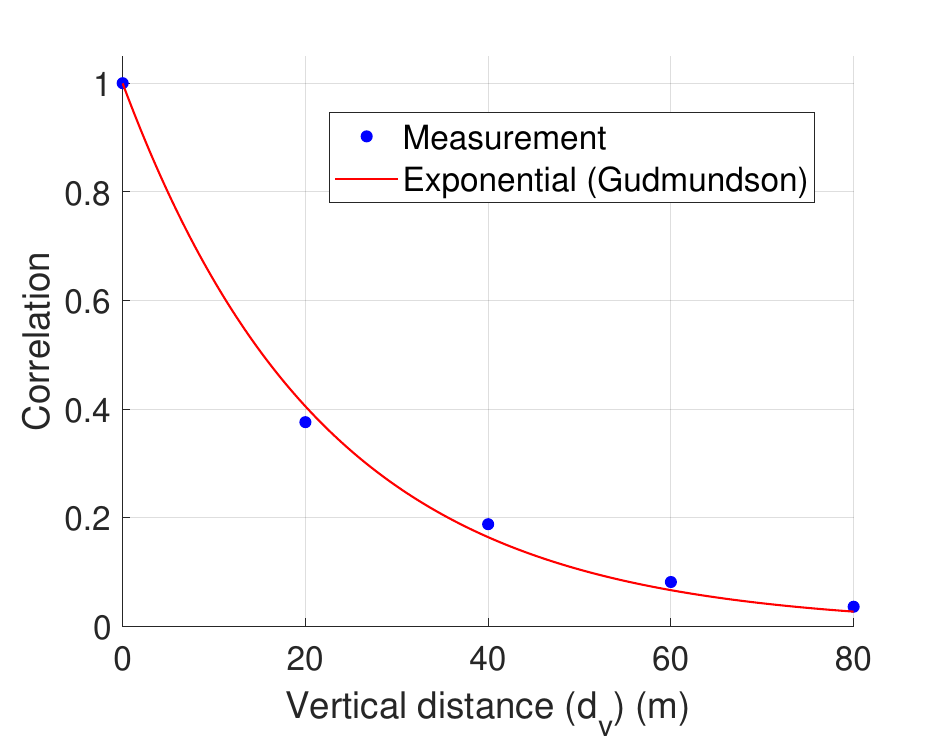}\label{fig:corr_dv}}
	\caption{Correlation of RSRP measurements with horizontal and vertical distance and curve fits with exponential and bi-expontential models.}
\end{figure}
% \begin{figure}[t]
% 	\centering
% 	\includegraphics[width=0.48\textwidth]{corr_dv.pdf}
% 	\caption{Correlation of RSRP measurements with vertical distance and curve fit with exponential model.}\label{fig:corr_dv}
% \end{figure}
We analyze the spatial correlation depending on the horizontal distance ($d_{\rm h}$) with a zero vertical distance ($d_{\rm v}$) by using the experiment dataset. Since our experiments fix the height of the drone for a specific flight, the vertical distance between the samples in the same flight is zero. The analysis of the correlation by the horizontal distance is performed by following steps:
\begin{enumerate}
  \item[i.] Calculate the correlation among all samples in a flight, excluding the samples during the take-off and landing periods.
  \item[ii.] Sort the correlation from step (i) according to the horizontal distance between the sample pairs. This will ensure that the correlations are arranged in increasing order based on the horizontal distance.
  \item[iii.] Average the correlations every 2~m. Start from the smallest horizontal distance and group the correlations within a 2~m interval. Calculate the average correlation for each interval. Repeat this process for subsequent 2~m intervals until covering all the correlations.
  \item[iv.] Perform steps (i)-(iii) iteratively for each height (30 m, 50 m, 70 m, 90 m, 110 m). Then, we have correlations for each individual height. 
  %Calculate the correlations separately for each flight at different heights.
  \item[v.] Average the correlation for every distance over all the heights. Take the correlations obtained in step (iv) for each height and distance, and compute the average correlation value across all heights for that specific distance.
\end{enumerate}
The correlation between two samples $w_{i}$, $w_{j}$ is calculated by
\begin{align}
    R_{i,j}=\frac{(w_{i}-\nu_i)(w_{j}-\nu_j)}{\sigma_{w,i}\sigma_{w,j}},
\end{align}
where $\nu$, $\sigma_{w}$ denote the mean and the standard deviation of the sample, which can be obtained from Table~II. The obtained correlation function and fitted curves are shown in Fig.~\ref{fig:corr_dh}. It is observed that the correlation is rapidly decayed as the horizontal distance increases. Although the correlation is generally modeled by an exponential function (also known as the Gudmundson model)~\cite{gudmundson1991correlation}, the bi-exponential model~\cite{liu2016investigation} fits better than the exponential model for our measurements, which is given as
\begin{align}\label{eq:corr_dh}
    R(d_{\rm h})=ae^{-b_1d_{\rm h}}+(1-a)e^{-b_2d_{\rm h}},
\end{align}
where $b_1$, $b_2$ are fitting parameters. We also observe that the correlation distance is 4.5~m when the correlation is 0.5.

\subsubsection{Vertical distance correlation}
\begin{table}[t]
\renewcommand{\arraystretch}{1.1}
\caption{Correlations between different UAV heights.}
\label{table:corr_dv}
\centering
\begin{tabular}{c||c|c|c|c|c}
\hline
& $30$~m & $50$~m & $70$~m & $90$~m & $110$~m\\
\hline
\hline
$30$~m & $1.003$ & $0.247$ & $0.080$ & $-0.024$ & $-0.040$\\
\hline
$50$~m & $0.247$ & $1.004$ & $0.214$ & $0.057$ & $-0.002$\\
\hline
$70$~m & $0.080$ & $0.214$ & $1.003$ & $0.307$ & $0.172$\\
\hline
$90$~m & $-0.024$ &	$0.057$ & $0.307$ &	$1.001$ & $0.409$\\
\hline
$110$~m & $-0.040$ & $-0.002$ &	$0.172$ & $0.409$ &	$1.012$\\
\hline
\end{tabular}
\vspace{-0.15in}
\end{table}
We calculate the vertical distance correlation with a zero horizontal distance from measurements which is opposite to the above subsection. Since the trajectory of the UAV for   flights at different heights is designed to be identical (see Fig.~\ref{fig:trajectory}), we can obtain samples of the same 2D location (latitude, longitude) with different vertical distances. For example, if we want to obtain 20 m vertical distance samples, we can use the dataset from the 30~m and 50~m UAV flights and pick two samples from any overlapped trajectory (one from the 30~m height, the other from the 50~m height). The analysis of the correlation by the vertical distance is conducted by following steps:
\begin{enumerate}
  \item[i.] Choose two different height measurements datasets, such as the datasets from the 30 m and 50 m UAV flights;
  \item[ii.] Remove data where the two trajectories are not fully overlapped, using a threshold of $d_{\rm h} > 3$~m. This ensures that we have data points with the same location across the trajectories;
  \item[iii.] Calculate the correlations between the two samples with the same location across the trajectories. Compute the correlation coefficient for each pair of samples and average them out. This will give you the correlation for a specific vertical distance (e.g., 20~m) between the two heights.
  \item[iv.] Repeat steps (i) to (iii) iteratively for pairs of measurements at different heights. For example, we can calculate correlations for the 50 m and 70 m flights, 70 m and 90 m flights, and so on.
\end{enumerate}
In step (ii), we exclude the samples that the trajectory is undesirably not overlapped by checking GPS readings. The correlations between different pairs of flights are listed in Table~III. We also present the obtained correlation function from Table~III and the fitted curve in Fig.~\ref{fig:corr_dv}. It is observed that the correlation function based on the vertical distance fits best with the exponential model, which is expressed as
\begin{align}\label{eq:corr_dv}
    R(d_{\rm v})=e^{-\frac{d_{\rm v}}{d_{\rm cor}}\text{ln}(2)},
\end{align}
where the correlation distance is given by $d_{\rm cor}=11.24$ m.

\subsubsection{3D distance correlation}
\begin{figure}[t]
	\centering
	\includegraphics[width=0.48\textwidth]{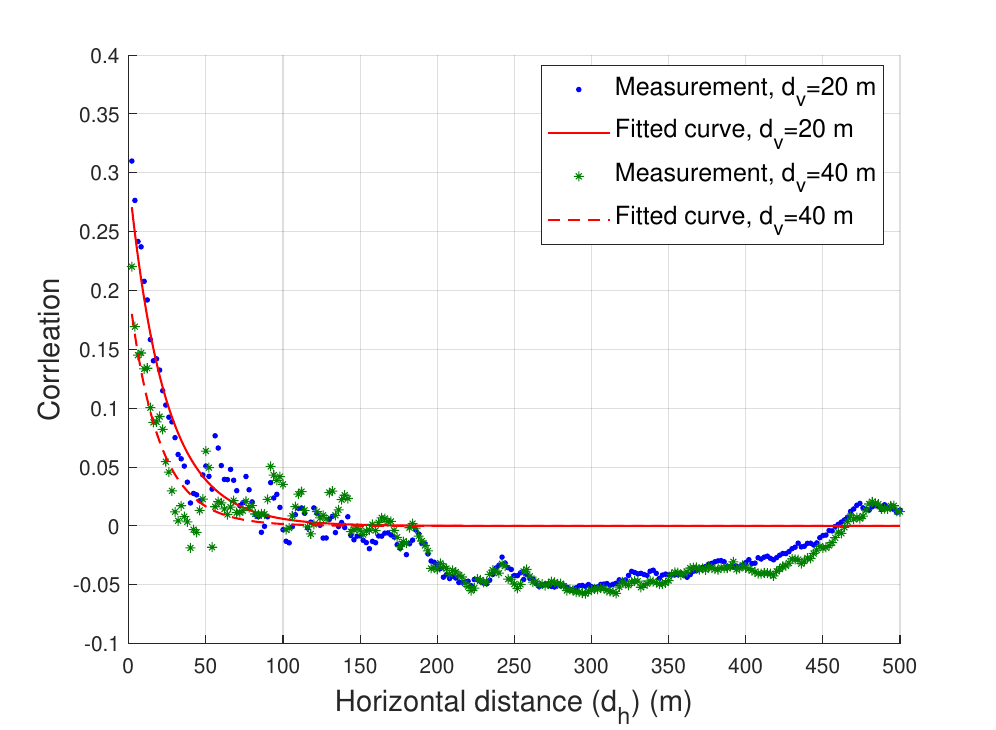}
	\caption{The 3D distance correlation and fitted curve by the proposed model where vertical distance ($d_{\rm v}$) = 20, 40 m.}\label{fig:corr_d3D}
\end{figure}
\begin{table}[t]
\renewcommand{\arraystretch}{1.1}
\caption{Fitting parameters $b_1$, $b_2$ in \eqref{eq:corr_d3D} depending on the vertical distance.}
\label{table:corr_b1b2}
\centering
\begin{tabular}{c|cc}
\hline
$d_{\rm v}$ & $b_1$ & $b_2$\\
\hline
0~m & 0.02815& 0.2474\\
20~m & 0.05988&	0.03574\\
40~m & 0.06998&	0.045\\
\hline
\hline
\end{tabular}
\vspace{-0.15in}
\end{table}
To analyze the correlation when both horizontal distance and vertical distance are considered, we can process the dataset obtained from flights at two different heights. By comparing the measurements from these flights, you can determine the correlation between two different 3D coordinate locations. The processing steps for obtaining correlation with 20 m vertical distance are as follows:
\begin{enumerate}
  \item[i.] Choose a pair of measurement datasets where the height difference is 20 m. For example, select the dataset from the 30 m height flight and the dataset from the 50 m height flight.
  \item[ii.] Calculate the correlation between a sample from one height (e.g., 30 m) and a sample from the other height (e.g., 50 m) across all the samples in the datasets.
  \item[iii.] Sort the correlation from step (ii) by the horizontal distance and average the correlations for every 2~m of horizontal distance.
  \item[iv.] Repeat steps (i) to step (iii) iteratively by different pairs of the measurement datasets of the height. For example, you can repeat the analysis with the dataset from the 50 m height flight and the dataset from the 70 m height flight.
\end{enumerate}
By performing this iterative analysis for different pairs of measurement datasets with varying height differences, we can obtain the correlation values that capture the relationship between joint horizontal and vertical distances. This analysis helps in understanding how the signal strength correlation varies with changes in both horizontal and vertical distances, providing insights into the spatial characteristics of the wireless channel.

The 3D distance correlation results with 20 m and 40 m vertical distances are shown in Fig.~\ref{fig:corr_d3D}. We model and fit the correlation of joint  horizontal and vertical distance by combining the correlation functions of the horizontal and the vertical distance in \eqref{eq:corr_dh}, \eqref{eq:corr_dv}. The proposed correlation model in 3D space is expressed as
\begin{align}\label{eq:corr_d3D}
    R(d_{\rm v}, d_{\rm h})=e^{-\frac{d_{\rm v}}{d_{\rm cor}}\text{ln}(2)}\left(ae^{-b_1d_{\rm h}}+(1-a)e^{-b_2d_{\rm h}}\right),
\end{align}
where $a=0.3$, and $b_1$, $b_2$ are tuning parameters. Note that when $d_{\rm h}=0$, the model is the same as \eqref{eq:corr_dh}, while when $d_{\rm v}=0$, the model is equivalent to \eqref{eq:corr_dv}. The fitted values of $b_1$, $b_2$ depending on the vertical distance ($d_{\rm v}$) are listed in Table~IV. 

\subsection{Analysis of Semi-variogram}
\begin{figure}[t]
	\centering
        \includegraphics[width=0.48\textwidth]{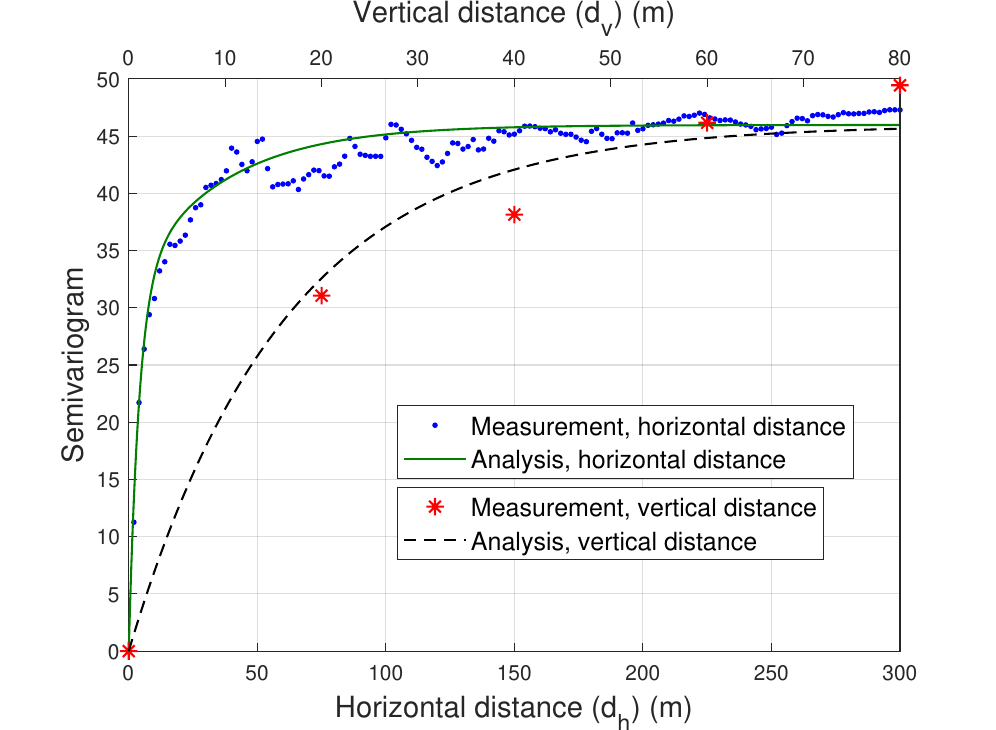}
	\caption{The measured semi-variogram and analysis by the correlation functions in~\eqref{eq:corr_dh} and ~\eqref{eq:corr_dv}, with respect to horizontal distance (bottom x-axis), and vertical distance (top x-axis).} \label{fig:semi_var}
\end{figure}
In Section \ref{sec:semi-var}, we introduce earlier the concept of semi-variogram in \eqref{eq:semi_var_def} and derive the relation to the correlation function in \eqref{eq:var2cor}. We analyze the semi-variogram by measurements results in Fig.~\ref{fig:semi_var} with respect to both the horizontal distance and vertical distance. The measurement results are directly obtained by the definition of the semi-variogram in \eqref{eq:semi_var_def} and the analysis results come from the correlation function in \eqref{eq:corr_d3D} which is then used in \eqref{eq:var2cor}. The measurements and our analysis from \eqref{eq:var2cor} are closely overlapped for both distance conditions.

\subsection{Performance Evaluation with Kriging}
In this subsection, we evaluate the 3D interpolation performance of the Kriging technique described in Section \ref{sec:Kriging_ana} using the measurement dataset. We adopt cross-validation-based root mean square error (RMSE) evaluation~\cite{9316892}, which compares the predicted RSRP with the measured RSRP to observe the error. In particular, the RMSE for performance evaluation can be expressed as
\begin{align}\label{eq:RMSE}
    \mathsf{RMSE}=\sqrt{\frac{1}{N_0}\sum^{N_0}_{i}\left(\hat{r}(l^{\rm uav}_{0,i})-r(l^{\rm uav}_{0,i})\right)^2},
\end{align}
where $N_0$ denotes the number of samples for prediction. In our evaluation, the 30~m height measurement samples are predicted by 30~m, 50~m, and 70~m height measurement datasets. The cross-validation-based evaluation is conducted by following steps:
\begin{enumerate}
  \item[i.] Randomly select $M$ samples from the measurement dataset to use for the prediction. These samples will serve as the training set.
  \item[ii.] Randomly select $N_0$ samples from the 30~m measurement dataset as the validation set for cross-validation.
  \item[iii.] Use the Kriging technique described in Section~\ref{sec:Kriging_ana} to predict the RSRP values for the $N_0$ validation samples based on the $M$ training samples.
  \item[iv.] Calculate RMSE between the predicted RSRP values and the actual measured RSRP values for the $N_0$ validation samples. The RMSE is calculated using \eqref{eq:RMSE}.
  \item[v.] Repeat steps (i) to (iv) iteratively for a large number of times, such as 10,000 iterations and calculate the median for the RMSE values obtained from the iterations. The median value represents the overall prediction performance of the Kriging technique.
\end{enumerate}

\begin{figure}[t]
	\centering
	\includegraphics[width=0.48\textwidth]{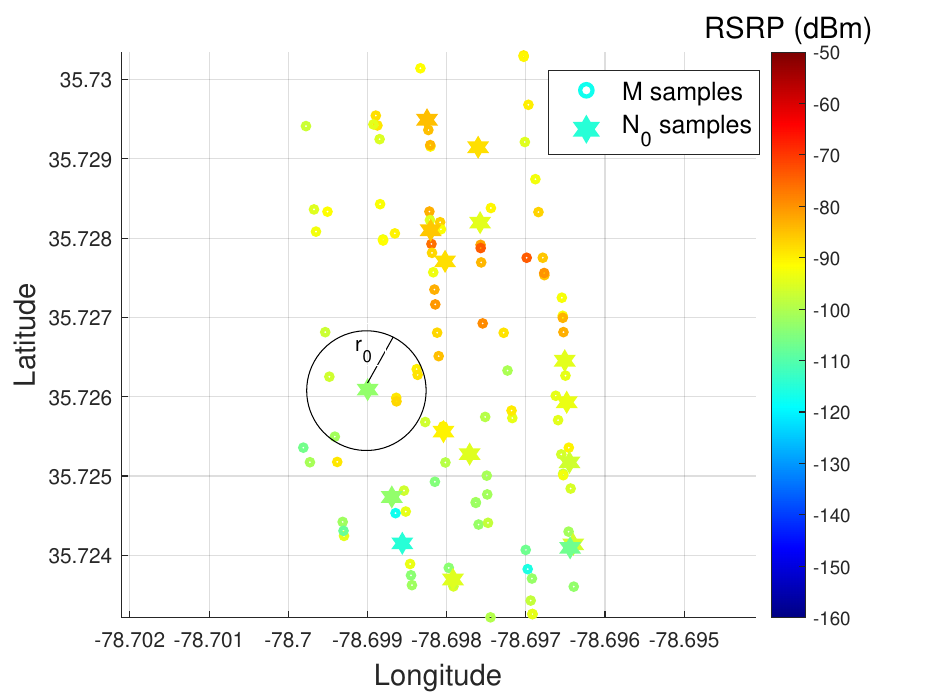}
	\caption{The snapshot of the randomly chosen samples for Kriging.}\label{fig:kriging_snapshot}
\end{figure}
\begin{figure}[t]
	\centering
	\subfloat[Prediction by 30 m height measurement.]{\includegraphics[width=0.44\textwidth]{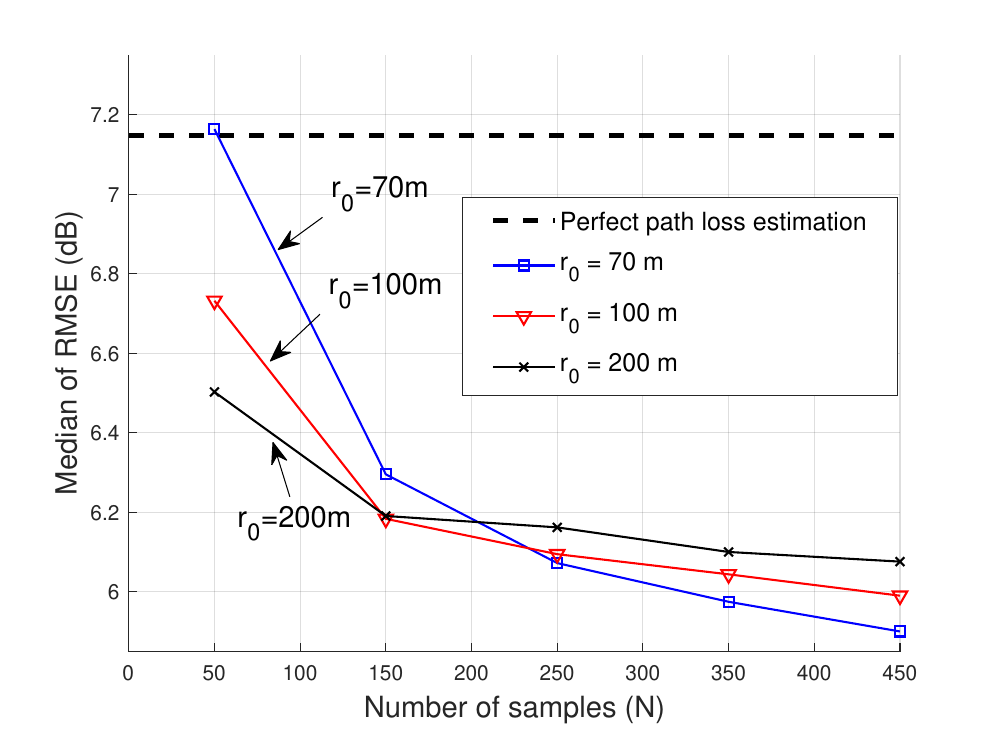}\label{fig:kriging_RMSE_30}}
	
	\subfloat[Prediction by 50 m height measurement.]{\includegraphics[width=0.44\textwidth]{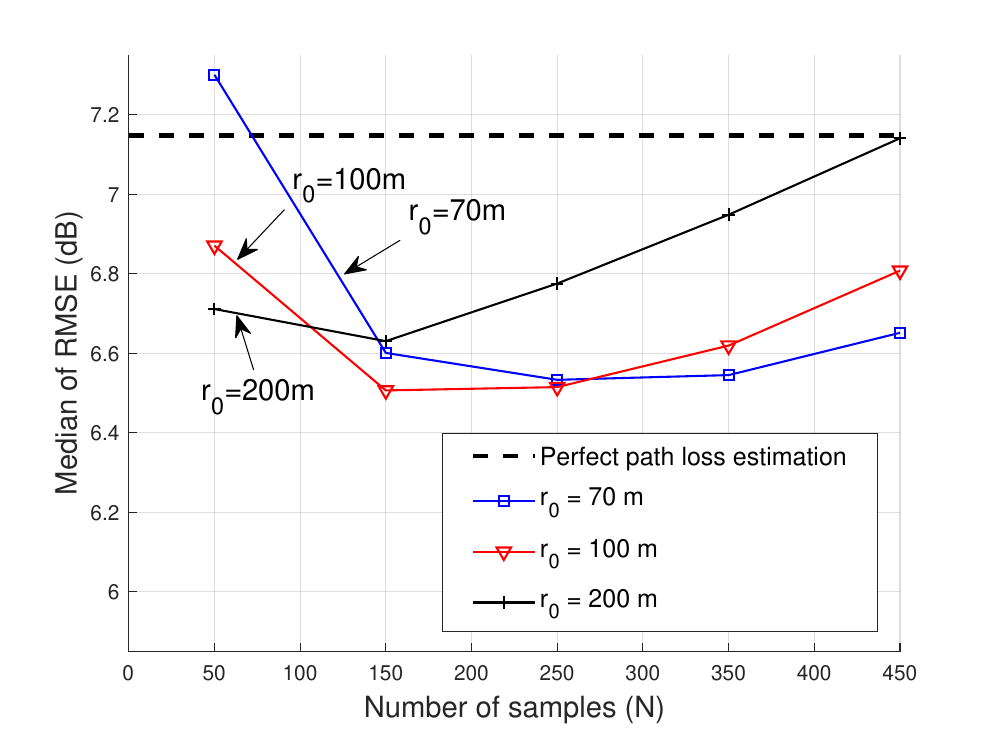}\label{fig:kriging_RMSE_50}}

	\subfloat[Prediction by 70 m height measurement.]{\includegraphics[width=0.44\textwidth]{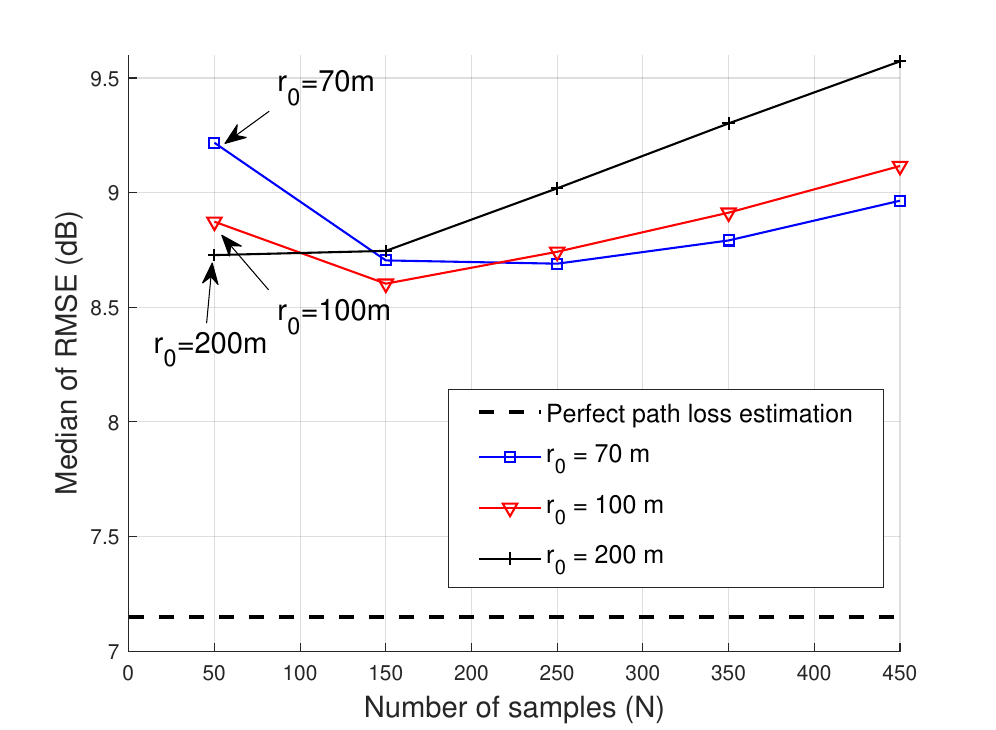}\label{fig:kriging_RMSE_70}}

	\caption{The RMSE of the prediction at 30~m UAV altitude from measurements at 30~m, 50~m, and 70~m UAV heights by Kriging, which is benchmarked by the perfect path loss estimation.}\label{fig:kriging_RMSE}
\end{figure}

In step (ii), after randomly selecting $M$ samples for prediction, exclude those samples from the dataset chosen for cross-validation. This ensures that the samples used for prediction are not used for validation. In addition, when we predict a sample by Kriging in step (iii), when predicting a sample using Kriging, consider only the nearby samples within a certain distance threshold ($r_0$). Limit the selection of neighboring samples to those within the $r_0$ radius circle around the target sample. These nearby samples will be used to predict the RSRP value for the target sample. The snapshot of the randomly chosen $M$ samples from 50~m height measurement and $N_0$ from 30~m height measurement is described in Fig.~\ref{fig:kriging_snapshot}. The figure depicts the radius circle $r_0$ within which nearby samples are used to predict the target sample. To provide a benchmark for comparison, we consider the perfect path loss-based 3D interpolation. In particular, we assume that the BS has perfect knowledge of the exact path loss and transmit power for all locations. This represents the ideal condition for prediction without utilizing spatial correlation. The RMSE by the perfect path loss estimation is equivalent to the standard deviation of the shadowing component from \eqref{eq:received} and \eqref{eq:RMSE} as follows:
\begin{align}\label{eq:RMSE_PLE}
    \mathsf{RMSE}_{\rm ple}=\sqrt{\mathbb{E}\left[\left(\hat{r}-r\right)^2\right]}=\sqrt{\mathbb{E}\left[w^2\right]}=\sigma_{w}.
\end{align}

In Fig.~\ref{fig:kriging_RMSE}, the RMSE performance of Kriging using measurements at different UAV altitudes is presented. The results show that the performance of Kriging varies depending on the altitude of the measurements used for prediction. When utilizing the 30 m and 50 m height measurement data for prediction, Kriging outperforms the perfect path loss estimation. This indicates that Kriging can leverage the spatial correlation present in the highly corrected data to achieve better prediction accuracy. However, in the case of 70~m height measurement, the perfect path loss estimation performs better than Kriging. This suggests that the correlation at a vertical distance of 60 m is too low to accurately predict using Kriging.

In Fig.~\ref{fig:kriging_RMSE_30}, it is observed that the RMSE generally decreases as the number of samples used for prediction ($N$) increases. However, when $N$ exceeds 250, the performance of Kriging with an $r_0$ value of 200~m is the worst among the three different r0 values considered. This indicates that while a larger number of samples can improve performance, adding low-correlated samples can degrade the prediction accuracy. It is important to strike a balance and choose an appropriate number of samples ($M$) and radius ($r_0$).

Furthermore, in Fig.~\ref{fig:kriging_RMSE_50}, the RMSE initially decreases and then increases for $r_0$ values of 70~m, 100~m, and 200~m. This suggests that if the correlation between samples is not sufficiently high, increasing the number of samples may not necessarily lead to improved performance. It highlights the importance of considering both the number of samples and the correlation when determining the optimal parameters for Kriging prediction.

In conclusion, the choice of the number of samples ($M$) and radius ($r_0$) is crucial for achieving accurate predictions using Kriging. Utilizing a larger number of highly correlated samples can improve performance, while including low-correlated samples or selecting an inappropriate radius can degrade the prediction accuracy.

\subsection{3D Interpolation by Kriging}
\begin{figure}[t]
	\centering
	\subfloat[Top view.]{\includegraphics[width=0.48\textwidth]{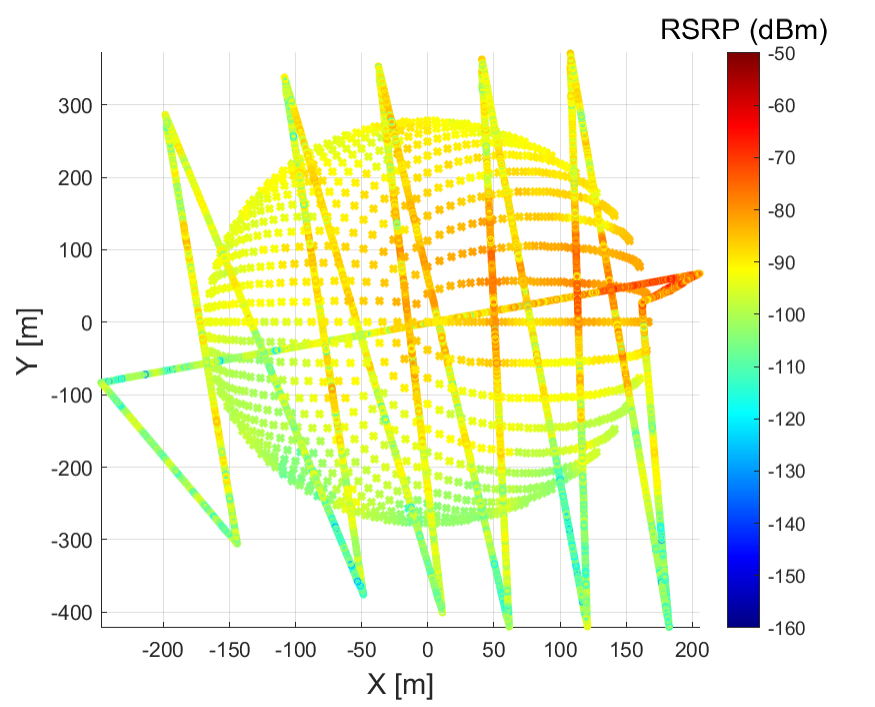}}
	
	\subfloat[3D view.]{\includegraphics[width=0.48\textwidth]{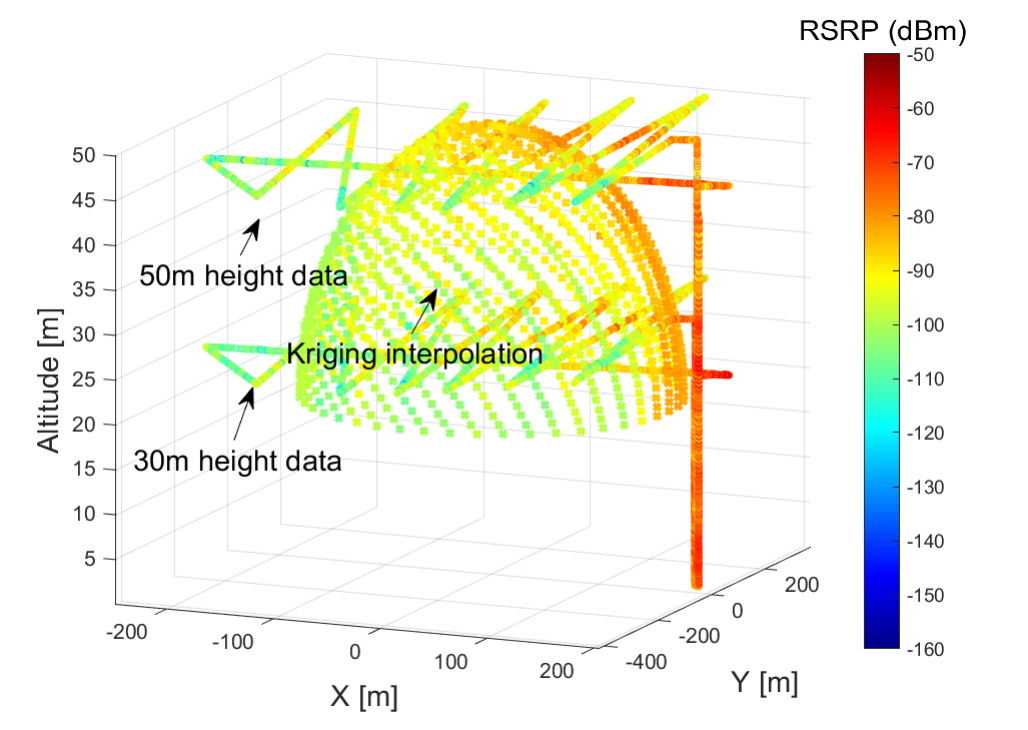}}
	\caption{The 3D radio map from Kriging interpolation using the measurements at 30~m and 50~m UAV altitudes.}\label{fig:kriging_3Dinterp}
\end{figure}

Fig.~\ref{fig:kriging_3Dinterp} displays the generated 3D radio map of RSRP using the Kriging interpolation technique with the available measurement data at 30 m and 50 m heights. The map provides a visual representation of the RSRP distribution in the 3D space. The dome shape of the 3D radio map provides valuable insights into monitoring the signal leakage in the three-dimensional volume of the RDZ. By examining the map, one can observe the spatial variations and signal strength levels within the monitored area. The dense 3D radio map obtained through Kriging interpolation enables efficient analysis and decision-making related to signal monitoring, interference management, and overall RF planning within the monitored area. In particular, a spectrum monitoring engine (SME) can estimate the received signal strength from each signal served within the RDZ on the surface of the dome. Subsequently, interference to sensitive receivers outside of the RDZ can be extrapolated, and if exceed a threshold, interfering signal services in the RDZ can take action (e.g. rescheduling to a different band or reducing power).

\section{Conclusion}\label{sec:conclusion}
In this paper, we introduce the RDZ concept which efficiently manages and controls the spectrum usage by monitoring the signal occupancy and leakage in a real-time fashion. To monitor the signal leakage from an area, we need to develop a radio map of signal power surrounding the area, which is more challenging when considering a 3D space. We propose a signal power interpolation method in the 3D volume that uses Kriging. The correlation model between two different 3D locations is designed and the semi-variogram is defined and analyzed. In addition, we study the proposed 3D Kriging interpolation using an experimental dataset provided by the NSF AERPAW platform. 
%A drone-mounted SDR and GPS receiver collects I/Q samples and we analyze the RSRP obtained by post-processing. 
We fit path loss and shadowing models to the RSRP measurements and study the performance of the Kriging interpolation technique for various scenarios. Our results show that significant gains are possible in received power estimation accuracy by utilizing the 3D correlation of the data when compared with using only a path loss based power estimation.  
%and compare it with the perfect path loss estimation.

% The authors also fit the path loss and shadowing components using the measurements, which are then used to evaluate the performance of the Kriging prediction. A comparison is made between the Kriging prediction and the perfect path loss estimation, which assumes the knowledge of exact path loss and transmit power for prediction without considering spatial correlation. By conducting this experimental validation and performance comparison, the paper provides insights into the effectiveness of the proposed 3D Kriging interpolation method for generating accurate radio maps and its potential applications in monitoring signal leakage and optimizing spectrum usage within the RDZ.
 
\bibliographystyle{IEEEtran} 
\bibliography{IEEEabrv,bibfile}
  
\end{document}